\documentclass[11pt,letterpaper]{article}
\usepackage[T1]{fontenc}
\usepackage{lmodern}
\usepackage{microtype}
\usepackage{amsmath,amssymb,amsthm,mathtools}
\usepackage[margin=1in]{geometry}
\usepackage{enumitem}
\usepackage[authoryear,round]{natbib}
\usepackage{graphicx}
\usepackage{booktabs}
\usepackage{tabularx}
\newcolumntype{Y}{>{\raggedright\arraybackslash}X}
\usepackage{float}
\usepackage{setspace}
\usepackage{hyperref}

\hypersetup{hidelinks}
\allowdisplaybreaks
\graphicspath{{figures/}}

\newtheorem{proposition}{Proposition}[section]
\newtheorem{lemma}[proposition]{Lemma}
\newtheorem{theorem}[proposition]{Theorem}
\newtheorem{corollary}[proposition]{Corollary}
\newtheorem{remark}[proposition]{Remark}

\newcommand{\E}{\mathbb E}
\newcommand{\Prob}{\mathbb P}
\newcommand{\Var}{\operatorname{Var}}

\title{\Large Designing Public Information for Strategic Advice\thanks{Refine.ink was used to check the paper for consistency and clarity.}}
\author{Vasudha Jain\thanks{Department of Economic Sciences, Indian Institute of Technology Kanpur. Email: \href{mailto:vasudhaj@iitk.ac.in}{\texttt{vasudhaj@iitk.ac.in}}.}
\and Jaideep Roy\thanks{Department of Economics, University of Bath. Email: \href{mailto:J.Roy@bath.ac.uk}{\texttt{J.Roy@bath.ac.uk}}.}}
\date{October 2026}

\begin{document}

\maketitle

\begin{abstract}
A principal chooses an action using advice from an expert who privately knows local conditions. Before receiving advice, she can commission a public study of an independent state, such as future demand, that changes the expected conflict between their preferred actions. We show that the communication protocol determines the optimal study. Under optimal mediation, the value of advice is convex in conflict, making full revelation optimal. Under commitment to an action rule, the principal may prefer no information, while informative optimal studies include a report certifying high conflict. Under unmediated cheap talk, changes in the number of equilibrium messages create upward kinks in the value of advice. Optimal studies avoid interior communication thresholds and can induce two distant communication regimes while skipping every regime in between, strictly outperforming both full revelation and no information. When the principal retains the raw data, selective public disclosure can improve advice without sacrificing forecasting precision. We discuss implications for fiscal forecasts and spending agencies.
\end{abstract}

\section{Introduction}\label{sec:introduction}

The information a firm makes public can change what its employees are willing to tell it. Consider a manufacturer choosing capacity for a new product line. Headquarters can commission a study of future demand. It also needs the regional manager's judgment about suppliers, crews and other local conditions that are hard to assess from a distance. Both parties see the study's report, the manager alone knows the local conditions, and headquarters must act on her advice before demand is realized. The study and the manager's advice concern different uncertainties, yet the study changes the conversation.

The reason is a conflict of interest that depends on demand. Suppose the manager gains from running a larger division. After an optimistic demand report, she wants to expand capacity by more than headquarters does. The report says nothing about supplier readiness, but it raises what the manager gains if headquarters believes readiness is high. Her account of local conditions becomes less credible exactly when the market outlook is good. A study can therefore help headquarters forecast demand while making it harder to learn about implementation. This paper asks how a decision maker should design public information when it also shapes the incentives of a later adviser.

\paragraph{What we do.} A principal commits to a public study of a payoff state that neither party observes. An expert privately knows a second, independent state. After seeing the study's report, she advises the principal, who then takes an irreversible action; only afterwards is the first state revealed. The two parties' preferred actions respond differently to the first state, so each report fixes the expected conflict under which the expert speaks. The same report also feeds the principal's own forecast.

\paragraph{An unusual instrument.} Many papers improve communication by limiting what the expert knows, or by keeping her preferences hidden from the receiver. Our study does neither. The expert knows her own state perfectly whatever the study says, and after every report the expected conflict is common knowledge. The study changes only how far apart the two parties are expected to be. And because Bayes' rule fixes the average conflict, the study cannot reduce it on average; it can only spread it across reports. The principal's instrument is a lottery over conflict.

\paragraph{The protocol determines the study.} Whether such a lottery is valuable depends on how the value of advice varies with the conflict, and that depends on how the conversation is organized. We compare three protocols in the uniform--quadratic model (Figure~\ref{fig:protocols}).
\begin{itemize}[leftmargin=2.2em]
\item Under optimal \emph{mediation} \citep{GoltsmanEtAl2009}, the loss from conflict rises steeply at first and then levels off. The value of advice is convex in the conflict, and the principal should reveal the conflict fully.
\item Under \emph{commitment} to an action rule \citep{Holmstrom1977}, the loss rises slowly at first, because a cap on discretion handles small conflicts well. The value is concave at low conflict and convex at high conflict. The principal should reveal nothing unless the conflict may be large, and then certify when it is (Section~\ref{sec:delegation}).
\item Under unmediated \emph{cheap talk} \citep{CrawfordSobel1982}, which is our main case, the value of advice equals the mediation value minus one ``scallop'' for each communication regime (Lemma~\ref{lem:scallops}). Cheap talk does exactly as well as mediation at one level of conflict in each regime and worse elsewhere. The convex trend pushes towards revealing; the scallops push towards blurring.
\end{itemize}

\paragraph{Main results under cheap talk.} The scallops produce three results. First, the principal never leaves the expert at the margin of losing a message (Proposition~\ref{prop:no-threshold}). Second, whenever the prior lies between two of the points where cheap talk matches mediation, saying nothing is strictly worse than some study (Proposition~\ref{prop:touch-points}). Third, the optimal study can place the expert in two communication regimes far apart, skipping every regime in between, while beating both full revelation and silence (Theorem~\ref{thm:global-skipping}). In our main example (Section~\ref{sec:main-example}), a fully revealing study would let the manager make five distinctions after weak-demand news and none after strong-demand news. The optimal study allows four distinctions after one report and two after the other; three is never used. A numerical map (Section~\ref{sec:how-common}) shows that blurring is common, occurring for about four priors in ten, while skipping is rarer, at about one case in twenty-five, with gains of up to about $10\%$ of the principal's loss.

\paragraph{A pure incentive device.} Blurring costs the principal forecasting precision only because the public report is her only demand information. If she keeps the raw data and publishes a coarser report, the expert's bias depends only on what is published, while the principal still acts on the raw data. Disclosure then serves communication alone, and the results survive (Section~\ref{sec:retained-information}). Under cheap talk, the instrument is valuable only because communication is discrete, and it costs the principal no information at all.

\paragraph{What to certify.} The protocols also differ in which news is most valuable. Under cheap talk, the most valuable news removes the conflict, because it restores full communication (Section~\ref{sec:zero-baseline}). Under commitment, the most valuable news shows that the conflict is large, because it tells the principal when to take discretion away. Under mediation, all news about the conflict is valuable.

\paragraph{A political application.} The same structure describes fiscal policy-making. A government sets the budget of a programme run by a spending agency or a regional government that knows local implementation conditions. The right budget also depends on future economic conditions, which an independent fiscal council forecasts in public. Agencies want larger budgets than the treasury, especially when the outlook is good. Our results say how precisely the official forecast should be published, whether the treasury should keep its own internal forecast, and how the answer changes under a fiscal rule that caps spending (Sections~\ref{sec:political-application} and~\ref{sec:fiscal-implications}).

\begin{figure}[t]
\centering
\includegraphics[width=0.98\textwidth]{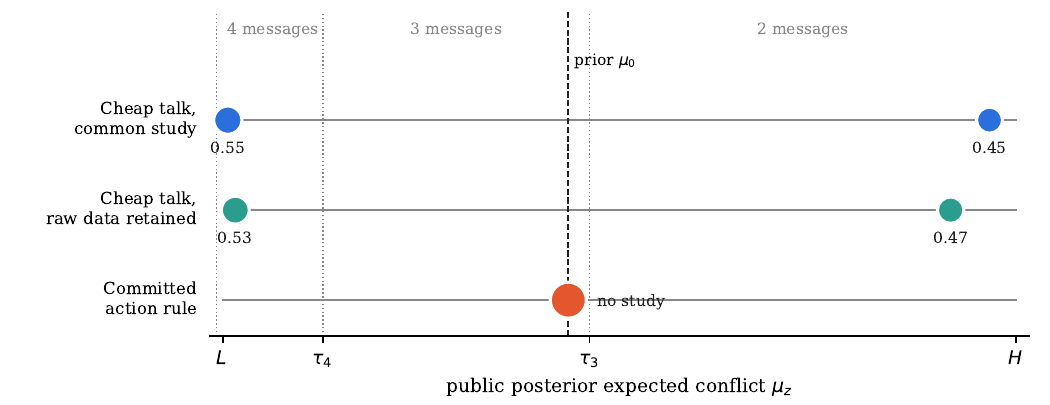}
\caption{\textbf{The optimal public study under three institutions} ($c=1/10$, $L=13/500$, $H=3/20$, $\mu_0=2/25$). The horizontal axis is the expected conflict implied by the public report; dotted lines mark where the expert loses a message. Dots show the reports of the optimal study, with their probabilities. Under cheap talk with a common study (Section~\ref{sec:positive-L}), the optimal study sends the expert to four or two messages and skips three. With retained raw data (Section~\ref{sec:retained-information}) the reports move slightly but the pattern is the same. With a committed action rule (Section~\ref{sec:delegation}) no study is run.}
\label{fig:overview}
\end{figure}

\subsection{Related literature}

Table~\ref{tab:literature} states the question each related literature asks and the question we ask.

\begin{table}[t]
\centering
\caption{Where the paper fits}
\label{tab:literature}
\small
\renewcommand{\arraystretch}{1.25}
\begin{tabularx}{\textwidth}{@{}>{\raggedright\arraybackslash}p{0.22\textwidth}YY@{}}
\toprule
Literature & Existing question & Our question \\
\midrule
Bayesian persuasion & Which beliefs should a designer induce in a receiver who then acts? & Which public beliefs should the receiver induce when they also set the value of a third party's advice? \\
Controlling the sender's knowledge & How much should the sender know about the state she reports? & The sender's knowledge is fixed; the study moves only her preferences. \\
Uncertain or hidden bias & Should the sender's privately known bias be disclosed? & The bias is common knowledge after every report; the receiver designs a lottery over it. \\
Public news and endogenous conflict & How do given public news or a chosen conflict affect communication? & How should the receiver design news when Bayes' rule fixes the average conflict? \\
Communication protocols & What is the best value of advice at a known bias? & How does the protocol shape the optimal information about the bias? \\
\bottomrule
\end{tabularx}
\end{table}

\emph{Information design.} The design step is a concavification \citep{KamenicaGentzkow2011}; see \citet{BergemannMorris2019}. Our contribution is to show what the object being concavified looks like when it is the value of strategic advice under different protocols, and what this implies for the design.

\emph{Controlling the sender's knowledge.} A large literature improves communication by limiting what the expert knows about the state she reports. \citet{FischerStocken2001} show that communication can be best when the sender is imperfectly informed. \citet{Ivanov2010} and \citet{AmbrusZhao2026} let the receiver garble the sender's signal; \citet{AmbrusZhao2026} show that nonmonotone garbling can help even in the uniform--quadratic model. \citet{LyuSuen2026} let an uninformed sender commit to an experiment whose outcome she then reports by cheap talk, and \citet{DeimenSzalay2019AER} let the expert choose what to learn. Related work includes \citet{Krahmer2021}, \citet{IvanovSam2022} and \citet{ArieliGeffnerTennenholtz2025}. In all these papers, the gain comes from the sender knowing less, so the receiver also learns less. In ours, the expert's knowledge of her state is untouched, and with retained data the principal gives up no information at all.

\emph{Uncertain or hidden bias.} When the sender privately knows her bias, keeping it uncertain or undisclosed can help the receiver \citep{MorganStocken2003,Li2004,LiMadarasz2008}. The mechanism there is pooling across privately informed sender types. In our model nobody has private information about the bias: after every report the expected conflict is common knowledge, and the receiver gains by randomizing \emph{which} commonly known conflict the expert faces.

\emph{Public news and endogenous conflict.} \citet{Chen2012} shows that more informative public news can hurt the receiver because the sender says less, and \citet{AgastyaBagChakraborty2014} show how uncertainty about a payoff state changes an expert's effective bias. Both take the news as given. \citet{AnticPersico2020} study communication when the conflict itself is chosen; our principal cannot choose the conflict, because Bayes' rule fixes its average. The receiver's own information also matters in \citet{MorenoDeBarreda2024} and \citet{MiyaharaSadakane2020}, but there she privately learns about the sender's state.\footnote{Other related settings allow an irreversible action to be delayed while public news arrives \citep{Frones2026}, or let the designer later communicate the information herself \citep{Jain2018,KreutzkampLou2025}.}

\emph{Communication protocols and political institutions.} \citet{GoltsmanEtAl2009} compare mediation, arbitration and negotiation at a known bias; the arbitration value coincides with optimal delegation \citep{Holmstrom1977,Holmstrom1984,MelumadShibano1991,AlonsoMatouschek2008}. We take these values as inputs and ask how each protocol shapes the optimal public information about the bias. The political application builds on the informational view of legislative committees \citep{GilliganKrehbiel1987}, on budget-maximizing bureaucracies \citep{Niskanen1971}, and on the role of independent fiscal councils \citep{BeetsmaDebrun2016}.

\paragraph{Roadmap.} Section~\ref{sec:model} sets up the model and its political interpretation. Section~\ref{sec:communication} describes cheap talk after a public report and compares it with mediation. Section~\ref{sec:positive-L} contains the main results under cheap talk and measures how common and how large the effects are. Section~\ref{sec:retained-information} lets the principal keep the raw data, and Section~\ref{sec:zero-baseline} treats the case in which weak demand removes the conflict. Section~\ref{sec:delegation} allows commitment to the action rule. Section~\ref{sec:protocols} puts the three protocols together, and Section~\ref{sec:conclusion} discusses the fiscal application and concludes. Proofs and further results are in the appendices.

\section{Model}\label{sec:model}

\subsection{Players, states and payoffs}

There are two players, a principal $P$ (headquarters) and an expert $E$ (the regional manager). The principal chooses an action $a\in\mathbb R$ (capacity). The payoff-relevant state has two independent components. The expert privately observes
\[
\theta\sim U[0,1],
\]
the quality of local implementation. A second state $s$ (future demand) is unknown to both players; it is independent of $\theta$ and has a finite second moment. Payoffs are
\begin{align}
u_P(a,\theta,s)&=-\left(a-\theta-w_Ps\right)^2, \label{eq:principal-payoff}\\
u_E(a,\theta,s)&=-\left(a-\theta-w_Es\right)^2, \label{eq:expert-payoff}
\end{align}
with $w_E>w_P\geq0$. Both players want more capacity when local conditions are good, but the expert's preferred capacity responds more strongly to demand: $w_E$ can capture divisional-growth incentives, sales-linked pay, or a manager who cares more about regional activity than headquarters does. Only the principal chooses the action. Let
\[
\delta\equiv w_E-w_P>0,
\qquad
B\equiv\delta s.
\]
If both states were known, the players' preferred actions would differ by $B$, the \emph{conflict}. In the binary benchmark, $B\in\{L,H\}$ with $0\leq L<H$. A positive $L$ means that the manager wants more capacity than headquarters even when demand is weak, as is natural for a product line that has already passed a viability test.

\subsection{The public study and timing}

The principal commits to a public study $\pi(z\mid s)$. Its report $z$ is seen by both players before the expert speaks; ``public'' means public within the firm. In the baseline, which we call the \emph{common study}, the report is the only information about demand that either player has when it matters: the expert sees $(\theta,z)$, and the principal acts on $(z,m)$, where $m$ is the expert's message. This can be implemented by hiring an independent research firm to publish only prespecified demand categories, or by not retaining the underlying data. Section~\ref{sec:retained-information} relaxes it.

\begin{figure}[t]
\centering
\includegraphics[width=0.98\textwidth]{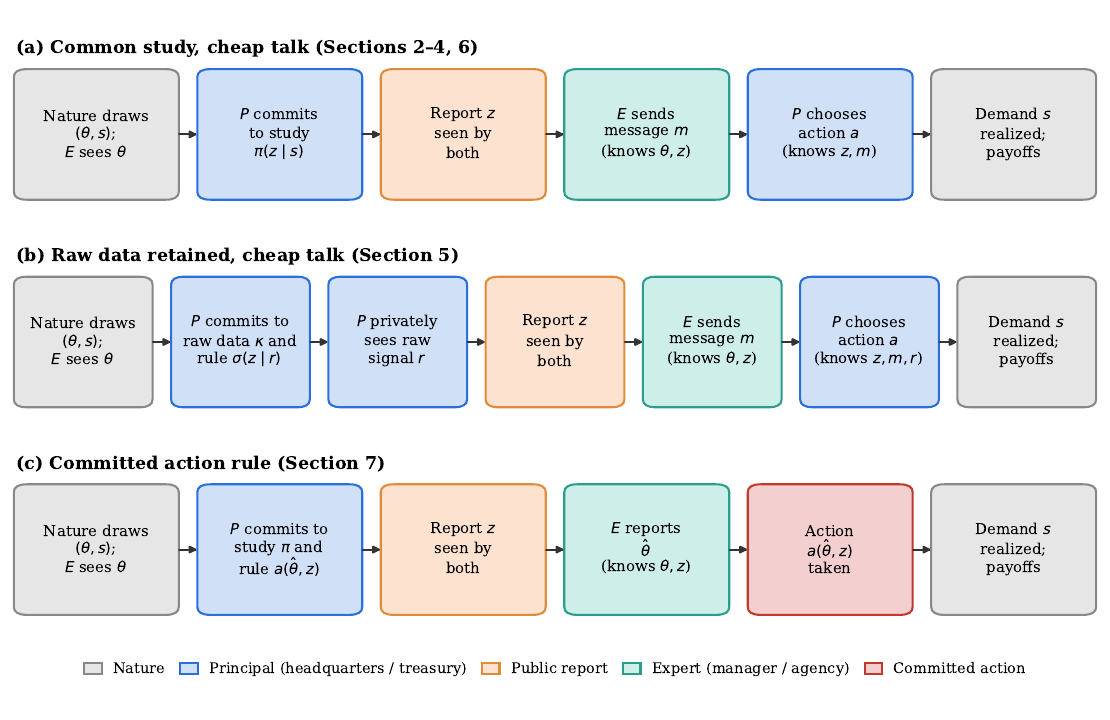}
\caption{\textbf{Timing under the three institutions.} Colours show who moves or what becomes known. Panel (a): the baseline, in which the public report is the principal's only demand information. Panel (b): the principal also keeps raw data $r$ and commits to how the public report is drawn from it. Panel (c): the principal commits in advance to the action as a function of the expert's report and the public report.}
\label{fig:timing}
\end{figure}

The timing is as follows.
\begin{enumerate}[label=\arabic*.,leftmargin=2.2em]
\item Nature draws $(\theta,s)$. The expert observes $\theta$.
\item The principal commits to the study $\pi$.
\item The report $z$ is publicly observed.
\item The expert sends a costless message $m$.
\item The principal chooses the irreversible action $a$.
\item Demand $s$ is realized and payoffs accrue.
\end{enumerate}
The principal commits to the study, not to an action rule (Section~\ref{sec:delegation} adds that). Because the action is taken before demand is known, the report matters in two ways: it improves the principal's forecast, and it changes the expert's incentives to communicate. Irreversibility rules out a second conversation after demand is known. Figure~\ref{fig:timing}(a) summarizes this timing; panels (b) and (c) show the two alternative institutions studied in Sections~\ref{sec:retained-information} and~\ref{sec:delegation}.

\subsection{The effective bias}

Fix a report $z$ and write
\[
e_z\equiv\E[s\mid z],
\qquad
\mu_z\equiv\E[B\mid z]=\delta e_z,
\qquad
q(m,z)\equiv\E[\theta\mid m,z].
\]
We call $\mu_z$ the \emph{expected conflict} after report $z$. The expert does not observe $s$, so her message carries no information about demand beyond $z$, and $\E[s\mid m,z]=e_z$.

\begin{lemma}[Effective bias]\label{lem:effective-bias}
After report $z$ and message $m$, the principal chooses
\begin{equation}
a^*(m,z)=q(m,z)+w_Pe_z. \label{eq:optimal-action}
\end{equation}
Given $(\theta,z)$, the part of the expert's expected payoff that depends on her message is
\begin{equation}
-\left(q(m,z)-\theta-\mu_z\right)^2. \label{eq:expert-reduced-payoff}
\end{equation}
The communication game after report $z$ is therefore the uniform--quadratic cheap-talk game with bias $\mu_z$.
\end{lemma}

\begin{proof}
The action \eqref{eq:optimal-action} is the conditional mean of the principal's ideal action $\theta+w_Ps$. Then $a^*-\theta-w_Es=q-\theta+w_Pe_z-w_Es$, whose last two terms have conditional mean $-\mu_z$ given $z$. Hence
\begin{equation}
\E\!\left[(a^*-\theta-w_Es)^2\mid\theta,m,z\right]=\left(q(m,z)-\theta-\mu_z\right)^2+w_E^2\Var(s\mid z), \label{eq:expert-loss-decomposition}
\end{equation}
and the variance term does not depend on the message.
\end{proof}

For the principal,
\begin{equation}
\E\!\left[(a^*-\theta-w_Ps)^2\mid m,z\right]=\Var(\theta\mid m,z)+w_P^2\Var(s\mid z). \label{eq:principal-loss-decomposition}
\end{equation}
The first term is the cost of imperfect advice; the second is the cost of not knowing demand. A public report affects both: it changes the expert's bias, and so the first term, and it changes the forecast error in the second term.

\subsection{A political application: fiscal forecasts and spending agencies}\label{sec:political-application}

The model has a direct political reading. The principal is a government, or its treasury, that sets the budget $a$ of a programme. The expert is the spending agency or regional government that runs the programme and knows how well it can be implemented locally ($\theta$). The right budget also depends on economic conditions over the budget period ($s$): growth, tax revenue, and the demand for the service. Nobody knows these when the budget is set. Many governments delegate the macroeconomic forecast to an independent fiscal council or statistics office, whose forecast is public and is seen by the treasury and the agencies alike \citep{BeetsmaDebrun2016}. That forecast is the public study $z$.

Spending agencies tend to want larger budgets than the treasury \citep{Niskanen1971}, and they want them especially when the economic outlook is good, because a buoyant economy makes a larger programme easier to defend and more visible. This is $w_E>w_P$, with $L>0$ when the agency wants more even in a downturn. Budget allocations are hard to reverse within the fiscal year, so the action is irreversible. The common-study architecture corresponds to a treasury that relies on the official forecast; the retained-data architecture of Section~\ref{sec:retained-information} corresponds to a treasury that keeps its own detailed internal forecast; and the committed rule of Section~\ref{sec:delegation} corresponds to a fiscal rule that caps the agency's budget as a function of the official forecast. A similar structure arises when a legislature relies on a committee's specialized knowledge \citep{GilliganKrehbiel1987} while an independent budget office publishes projections that shape how far the committee's preferences diverge from the floor's. Section~\ref{sec:fiscal-implications} collects the implications.

\section{Cheap Talk After a Public Report}\label{sec:communication}

\subsection{How many messages can the expert send?}

Fix expected conflict $\mu\geq0$. As in \citet{CrawfordSobel1982}, an equilibrium with $j$ messages partitions $[0,1]$ into intervals $0=t_0<t_1<\cdots<t_j=1$. Types in $[t_{i-1},t_i]$ send the same message, and the principal's action (net of $w_Pe_z$) is the midpoint $q_i=(t_{i-1}+t_i)/2$. A boundary type must be indifferent between the two neighbouring actions, $t_i+\mu=(q_i+q_{i+1})/2$, which gives
\begin{equation}
t_{i+1}=2t_i-t_{i-1}+4\mu. \label{eq:boundary-recurrence}
\end{equation}
Each interval is $4\mu$ longer than the one before. Since the lengths sum to one,
\begin{equation}
\ell_i^{(j)}=\frac1j+2\mu(2i-j-1),\qquad i=1,\ldots,j. \label{eq:cell-lengths}
\end{equation}
The first interval is the shortest, so a $j$-message equilibrium exists if and only if $\ell_1^{(j)}>0$, that is,
\begin{equation}
\mu<\tau_j\equiv\frac1{2j(j-1)},\qquad j\geq2. \label{eq:thresholds}
\end{equation}
The thresholds $\tau_2=1/4$, $\tau_3=1/12$, $\tau_4=1/24$, \ldots\ fall towards zero. As expected conflict rises past $\tau_j$, the finest equilibrium loses its $j$-th message. Above $1/4$ only babbling remains. At $\mu=0$ the expert can reveal $\theta$ fully.

\subsection{The value of advice}

In an interval of length $\ell$, the principal's mean squared error about $\theta$ is $\ell^2/12$. Weighting by the interval lengths and using \eqref{eq:cell-lengths}, the value of advice in the $j$-message equilibrium is
\begin{equation}
V_j(\mu)=-\frac1{12j^2}-\frac{j^2-1}{3}\mu^2,\qquad j\geq2, \label{eq:Vj}
\end{equation}
and babbling gives
\begin{equation}
V_1(\mu)=-\frac1{12}. \label{eq:V1}
\end{equation}
By Lemma~\ref{lem:effective-bias}, both players' payoffs differ from $V_j$ only by terms that do not depend on the number of messages, so both prefer more messages:
\begin{equation}
V_{j+1}(\mu)-V_j(\mu)=\frac{2j+1}{3}\left(\tau_{j+1}^2-\mu^2\right)>0\quad\text{whenever }\mu<\tau_{j+1}. \label{eq:adjacent-ranking}
\end{equation}

\paragraph{Which equilibrium is played.} Cheap talk has many equilibria, including babbling. We assume that the finest equilibrium is played at every public report that the optimal study actually uses. This is what the No Incentive to Separate criterion of \citet{CKS2008} selects: the $j$-message equilibrium passes the criterion if and only if
\begin{equation}
\tau_{j+1}\leq\mu\leq\tau_j, \label{eq:NITS-condition}
\end{equation}
so for
\begin{equation}
\mu\in I_j\equiv[\tau_{j+1},\tau_j) \label{eq:regime-interval}
\end{equation}
it selects $j$ messages.\footnote{The lowest type's equilibrium loss is $(\frac1{2j}-j\mu)^2$; separating to the action $0$ would cost her $\mu^2$. The criterion compares the two, which gives \eqref{eq:NITS-condition}.} We call $I_j$ the \emph{$j$-message regime}. Three facts make the finest equilibrium a natural prediction here. First, at every report both players rank all partition equilibria in the same order, by \eqref{eq:adjacent-ranking} and Lemma~\ref{lem:effective-bias}, so the finest equilibrium is the one \emph{both} prefer; neither has a reason to coordinate on a coarser one. Second, the same outcome is selected by the No Incentive to Separate criterion, by either player's preferred equilibrium, and by ex ante Pareto efficiency. Third, our design results need this selection only at the reports the optimal study actually uses; elsewhere any partition equilibrium may be played (Remark~\ref{rem:selection-robustness}). What cannot be ruled out is babbling, which is an equilibrium after every report; our results do not survive if babbling is played at the reports the study uses. \citet{AnticPersico2020} show how some conclusions about endogenous conflict can be reached without selecting an equilibrium; doing so here is left open. All equilibrium statements concern monotone pure-strategy partition equilibria.

\begin{figure}[t]
\centering
\includegraphics[width=0.94\textwidth]{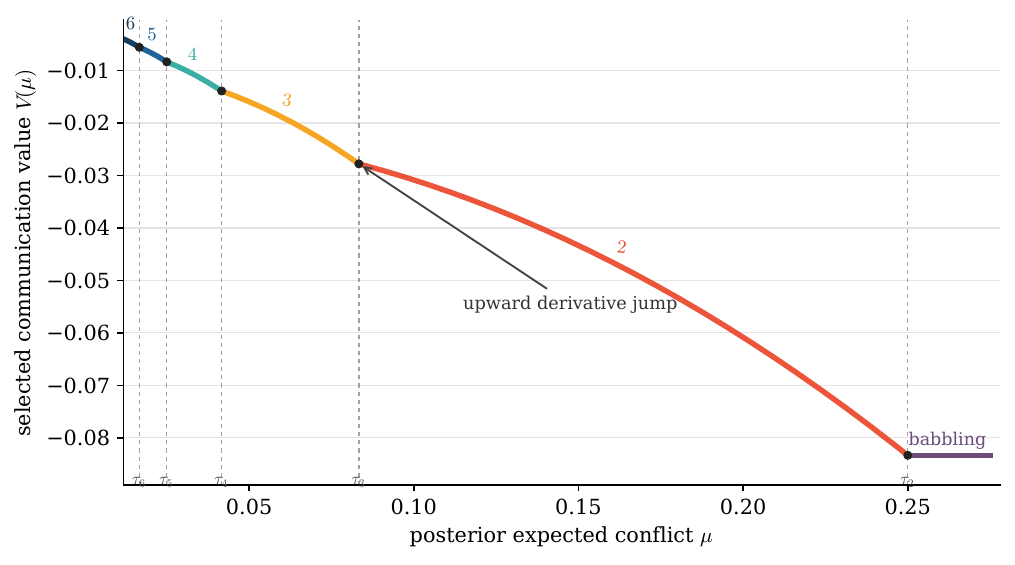}
\caption{\textbf{The value of advice as a function of expected conflict.} Numerals give the number of messages in the selected equilibrium. The value is concave within each regime, flat under babbling, and continuous at each threshold $\tau_j$, but its slope jumps up when a message disappears. Adding the smooth forecast term $c\mu^2$ of Section~\ref{sec:design-objective} changes the curvature within regimes but not the kinks.}
\label{fig:communication-capacity}
\end{figure}

The selected value of advice is
\begin{equation}
V(\mu)=V_j(\mu)\quad\text{for }\mu\in I_j,\qquad V(\mu)=-\tfrac1{12}\ \text{for }\mu\geq\tfrac14,\qquad V(0)=0. \label{eq:selected-V}
\end{equation}
It is continuous and strictly concave within each regime, since $V_j''=-2(j^2-1)/3<0$. At each threshold, however, its slope jumps \emph{up}:
\begin{equation}
V'_+(\tau_j)-V'_-(\tau_j)=\frac{2j-1}{3j(j-1)}>0. \label{eq:derivative-jump}
\end{equation}
Just below $\tau_j$ the expert still has $j$ messages and each extra unit of conflict squeezes all of them; just above it, she has one fewer message, and the remaining ones are squeezed less. Figure~\ref{fig:communication-capacity} shows the result: a curve that is concave in pieces, with a convex kink wherever a message disappears. These kinks are the source of everything that follows.

\subsection{The principal's design problem}\label{sec:design-objective}

Combining \eqref{eq:principal-loss-decomposition} with the selected equilibrium, the principal's ex ante payoff from a study is
\begin{equation}
\E_z\!\left[V(\mu_z)-w_P^2\Var(s\mid z)\right]=\E_z\!\left[V(\mu_z)-c\Var(B\mid z)\right],
\qquad
c\equiv\left(\frac{w_P}{w_E-w_P}\right)^2. \label{eq:general-design-objective}
\end{equation}
The parameter $c$ measures how much the principal's own decision depends on demand, relative to the conflict. With binary conflict, $\Var(B\mid z)=(\mu_z-L)(H-\mu_z)$, so
\begin{equation}
\Var(B\mid z)=-\mu_z^2+(L+H)\mu_z-LH. \label{eq:binary-variance}
\end{equation}
Since $\E[\mu_z]=\mu_0\equiv\E[B]$ for every study, the principal's payoff equals $\E[W(\mu_z)]$ up to the constant $-c(L+H)\mu_0+cLH$, where
\begin{equation}
W(\mu)\equiv V(\mu)+c\mu^2. \label{eq:W-definition}
\end{equation}
The two parts of $W$ have clear meanings. $V(\mu)$ is the value of advice. The term $c\mu^2$ is the value of the forecast: by the law of total variance, a study reduces expected forecast error by $c\Var(\mu_z)$, which is largest under full revelation. On a $j$-message regime, writing $t\equiv1+3c$,
\begin{equation}
W_j(\mu)=-\frac1{12j^2}-\frac{j^2-t}{3}\mu^2, \label{eq:Wj-t}
\end{equation}
which is strictly concave if $t<j^2$; under babbling, $W_1(\mu)=-\frac1{12}+c\mu^2$. The forecast term does not change the kinks in \eqref{eq:derivative-jump}.

The principal's problem is therefore to choose a distribution of posterior means $\mu_z$ with mean $\mu_0$ that maximizes $\E[W(\mu_z)]$; that is, to find the concave envelope of $W$ \citep{KamenicaGentzkow2011}. Spreading beliefs pays where $W$ is convex, which here means around the kinks.

\subsection{Cheap talk as mediation with scallops}\label{sec:scallops}

To see what the kinks do, compare cheap talk with a benchmark. Suppose that after the report, the expert speaks to a neutral mediator who then recommends an action, and that the mediator is designed optimally. \citet{GoltsmanEtAl2009} show that in the uniform--quadratic model with a known bias $\mu$, the best the receiver can obtain this way is
\begin{equation}
V^M(\mu)=-\frac{\mu(1-\mu)}{3}\quad\text{for }\mu\leq\tfrac12,\qquad V^M(\mu)=-\frac1{12}\quad\text{for }\mu\geq\tfrac12. \label{eq:VM}
\end{equation}
This value is convex: the first units of conflict are the most costly, and the cost levels off as conflict grows.

\begin{lemma}[Cheap talk is mediation minus scallops]\label{lem:scallops}
For every $j\geq1$ and every $\mu$ in the $j$-message regime (with $I_1=[1/4,1/2]$ for babbling),
\begin{equation}
V(\mu)=V^M(\mu)-\frac{(2j^2\mu-1)^2}{12j^2}. \label{eq:scallop-identity}
\end{equation}
Hence $V\leq V^M$, with equality exactly at $\mu=0$, at $\mu\geq1/2$, and at the points
\begin{equation}
p_j\equiv\frac1{2j^2},\qquad j=1,2,\ldots, \label{eq:touch-points}
\end{equation}
one in each regime ($p_1=1/2$, $p_2=1/8$, $p_3=1/18$, \ldots).
\end{lemma}

\begin{proof}
For $j\geq2$, expanding the right side of \eqref{eq:scallop-identity} gives $-\frac\mu3+\frac{\mu^2}3-\frac{j^2\mu^2}3+\frac\mu3-\frac1{12j^2}=V_j(\mu)$. For $j=1$ and $\mu\in[1/4,1/2]$, $V^M(\mu)-\frac{(2\mu-1)^2}{12}=-\frac1{12}=V_1$. The point $p_j$ lies strictly inside $I_j$ because $2j(j-1)<2j^2<2j(j+1)$.
\end{proof}

\begin{figure}[t]
\centering
\includegraphics[width=0.98\textwidth]{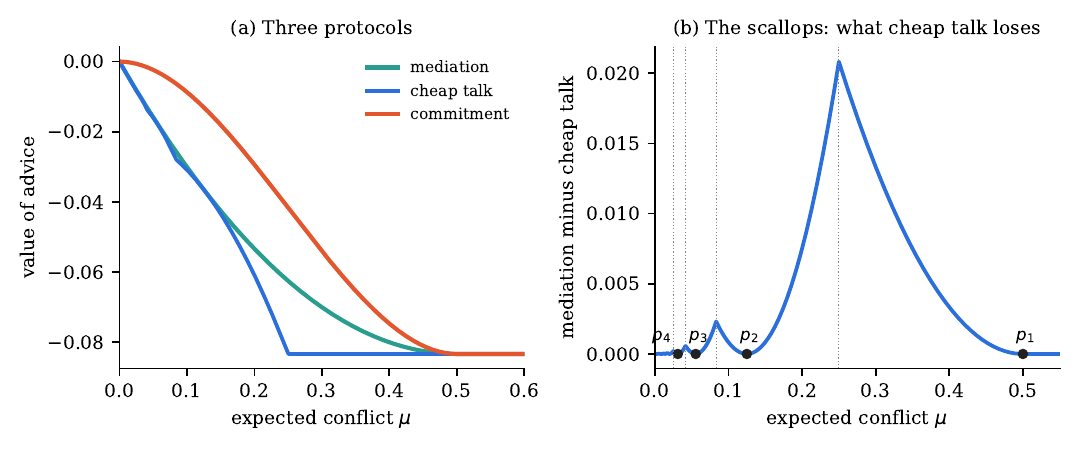}
\caption{\textbf{Three ways to communicate.} Panel (a): the value of advice as a function of expected conflict under optimal mediation (convex), unmediated cheap talk (finest equilibrium), and commitment to an action rule (concave, then convex). Panel (b): the gap between mediation and cheap talk. It is a row of scallops, one per communication regime, and vanishes at the touch points $p_j=1/(2j^2)$; dotted lines mark the thresholds $\tau_j$, where consecutive scallops meet.}
\label{fig:protocols}
\end{figure}

Figure~\ref{fig:protocols} shows the picture. Cheap talk follows a convex curve from below and touches it once in each regime. At the touch point $p_j$, the finest $j$-message partition is as good as optimal mediation; away from it, cheap talk loses a scallop. The kinks of Section~\ref{sec:communication} are the joints between consecutive scallops. Panel (a) also shows the value under commitment, which Section~\ref{sec:delegation} derives. The three curves have different shapes, and Section~\ref{sec:protocols} shows that the shape determines the optimal study.

\section{Blurring the Forecast to Keep Advice Alive}\label{sec:positive-L}

Throughout this section the conflict is positive in both states, $0<L<H$. Let $p\equiv\Prob(B=H)\in(0,1)$, so $\mu_0=(1-p)L+pH$.

\subsection{Which studies are feasible}

A study induces a distribution of posterior means on $[L,H]$ whose average is the prior,
\begin{equation}
\E[\mu_z]=\mu_0. \label{eq:Bayes-plausibility}
\end{equation}
With a binary state, every such distribution can be produced by some study. If the target distribution puts probability $\lambda_r$ on posterior mean $\mu_r$, let $q_r=(\mu_r-L)/(H-L)$ be the implied posterior probability of the high state. The study that sends report $r$ with probabilities
\begin{equation}
\pi(r\mid H)=\frac{\lambda_rq_r}{p},
\qquad
\pi(r\mid L)=\frac{\lambda_r(1-q_r)}{1-p} \label{eq:binary-implementation}
\end{equation}
does the job. The principal can therefore choose any distribution of posterior means with mean $\mu_0$, and her best payoff is the concave envelope of $W$ at $\mu_0$.

\subsection{The basic principle: never at the margin of losing a message}

\begin{proposition}[Never at the margin]\label{prop:no-threshold}
Let $\tau_j\in(L,H)$ be a communication threshold.
\begin{enumerate}[label=\textup{(\roman*)},leftmargin=2.2em]
\item No optimal study ever produces the posterior $\tau_j$.
\item There is an open interval of priors around $\tau_j$ for which revealing nothing is strictly worse than some informative study.
\end{enumerate}
Both statements hold for any continuation value, not only the Crawford--Sobel one, at every point where its slope jumps up.
\end{proposition}

\begin{proof}
(i) Let $\eta$ be an optimal distribution of posteriors and $T$ a straight line that supports the concave envelope of $W$ at $\mu_0$. Then $T\geq W$ everywhere and $\E_\eta[W(\mu_z)]=T(\mu_0)=\E_\eta[T(\mu_z)]$, so $W=T$ at every posterior the study uses. A line can touch $W$ from above at $\tau_j$ only if its slope lies between the left slope and the right slope of $W$ there, and so is at most the left slope and at least the right slope. At $\tau_j$ the right slope is strictly larger than the left slope \eqref{eq:derivative-jump}, so no such line exists. (ii) At $\mu_0=\tau_j$, revealing nothing would use the posterior $\tau_j$, which (i) rules out; so the concave envelope lies strictly above $W$ at $\tau_j$. Both functions are continuous, so the gap stays positive for nearby priors.
\end{proof}

In words, the principal never leaves the expert at the exact margin where she is about to lose a message. Near such a margin the expert's advice is about to fall in quality, and it is cheaper to move her away from it in both directions, some of the time towards more messages and some of the time towards fewer, than to keep her there. This is the general principle behind our results. It needs only that the value of advice falls in steps, which is a property of cheap talk with a finite number of messages in many models, not only the uniform--quadratic one. What the uniform--quadratic model adds is a sharp answer to the next question: how far apart the two sides can be.

The scallop picture of Lemma~\ref{lem:scallops} gives a second, global reason to inform.

\begin{proposition}[Information helps between touch points]\label{prop:touch-points}
Suppose $[L,H]$ contains two touch points $p_n<p_m$ ($n>m\geq1$), and the prior satisfies $p_n<\mu_0<p_m$. Then the study whose two reports lead to $p_n$ and $p_m$ gives the principal strictly more than revealing nothing.
\end{proposition}

\begin{proof}
Let $W^M\equiv V^M+c\mu^2$. On $[0,1/2]$, $W^M(\mu)=-\frac\mu3+\frac{1+3c}{3}\mu^2$ is strictly convex. By Lemma~\ref{lem:scallops}, $W=W^M$ at $p_n$ and $p_m$, and $W\leq W^M$ everywhere. If $\lambda$ is the weight on $p_n$ that gives mean $\mu_0$, then
\[
\lambda W(p_n)+(1-\lambda)W(p_m)=\lambda W^M(p_n)+(1-\lambda)W^M(p_m)>W^M(\mu_0)\geq W(\mu_0).\qedhere
\]
\end{proof}

The logic is simple. At a touch point, cheap talk is as good as the best mediator. A study that sends the expert to touch points therefore secures the mediated value on average, and because the mediated value is convex, that average beats leaving her at the prior. In the main example below, the prior $\mu_0=2/25$ lies between $p_3=1/18$ and $p_2=1/8$, so silence cannot be optimal. Proposition~\ref{prop:no-threshold} says where the optimal study does \emph{not} go; Proposition~\ref{prop:touch-points} says that it goes \emph{somewhere} whenever the support spans two touch points. The theorem below says where.

\subsection{A line that touches two regimes}

The candidate optimum touches $W$ in two regimes, $n$ messages at a low posterior and $m<n$ messages at a high one. Fix integers $n>m\geq2$ and assume
\begin{equation}
t<m^2, \label{eq:t-below-m}
\end{equation}
so that every branch $W_k$ with $m\leq k\leq n$ is strictly concave. Define
\begin{equation}
x_{n,m}\equiv\frac1{2mn}\sqrt{\frac{m^2-t}{n^2-t}}, \label{eq:xnm}
\end{equation}
\begin{equation}
y_{n,m}\equiv\frac1{2mn}\sqrt{\frac{n^2-t}{m^2-t}}, \label{eq:ynm}
\end{equation}
and the straight line
\begin{equation}
T_{n,m}(\mu)\equiv-\frac{t}{12m^2n^2}-\frac{\sqrt{(n^2-t)(m^2-t)}}{3mn}\,\mu. \label{eq:Tnm}
\end{equation}

\begin{lemma}[One line, two regimes]\label{lem:common-tangent}\label{lem:intermediate-domination}
Assume \eqref{eq:t-below-m}.
\begin{enumerate}[label=\textup{(\roman*)},leftmargin=2.2em]
\item The line $T_{n,m}$ is tangent to the $n$-message branch $W_n$ at $x_{n,m}$ and to the $m$-message branch $W_m$ at $y_{n,m}$:
\begin{equation}
T_{n,m}(\mu)-W_n(\mu)=\frac{n^2-t}{3}(\mu-x_{n,m})^2,
\qquad
T_{n,m}(\mu)-W_m(\mu)=\frac{m^2-t}{3}(\mu-y_{n,m})^2. \label{eq:n-contact-square}
\end{equation}
\item The line lies strictly above every intermediate branch: for $m<k<n$,
\begin{equation}
\min_{\mu}\{T_{n,m}(\mu)-W_k(\mu)\}=\frac{t(k^2-m^2)(n^2-k^2)}{12k^2m^2n^2(k^2-t)}>0. \label{eq:intermediate-minimum}
\end{equation}
\item The contact $x_{n,m}$ lies in the $n$-message regime and $y_{n,m}$ lies in the $m$-message regime if and only if
\begin{equation}
t<\frac{m^2(2n+1)}{(n+1)^2-m^2} \label{eq:x-lower-condition}
\end{equation}
and
\begin{equation}
t<\frac{n^2(2m-1)}{n^2-(m-1)^2}. \label{eq:y-upper-condition}
\end{equation}
\end{enumerate}
\end{lemma}

Parts (i)--(ii) are algebra about two quadratic branches and their common tangent; part (iii) checks that the tangency points lie in the right regimes. Appendix~\ref{app:tangent-algebra} gives the details.

\subsection{The main result}

\begin{theorem}[Optimal blurring across distinct regimes]\label{thm:global-skipping}
Fix $n>m\geq2$ with $t<m^2$, and suppose the support satisfies
\begin{equation}
\tau_{n+1}<L<x_{n,m}<\tau_n \label{eq:lower-support-condition}
\end{equation}
and
\begin{equation}
\tau_{m+1}<y_{n,m}<H<\tau_m. \label{eq:upper-support-condition}
\end{equation}
Then for every prior $\mu_0\in(x_{n,m},y_{n,m})$, the unique optimal study has two reports. One leads to expected conflict $x_{n,m}$ and $n$ messages, the other to $y_{n,m}$ and $m$ messages, with probabilities
\begin{equation}
\Prob(\mu_z=x_{n,m})=\frac{y_{n,m}-\mu_0}{y_{n,m}-x_{n,m}}, \label{eq:lambda-x}
\end{equation}
\begin{equation}
\Prob(\mu_z=y_{n,m})=\frac{\mu_0-x_{n,m}}{y_{n,m}-x_{n,m}}. \label{eq:lambda-y}
\end{equation}
The study never induces any regime between $m$ and $n$; when $n\geq m+2$ it skips $n-m-1$ regimes. It strictly beats both no information and full revelation.
\end{theorem}

\begin{proof}
On $[L,H]$ the only regimes present are $n,n-1,\ldots,m$. By Lemma~\ref{lem:common-tangent}, $T_{n,m}$ lies weakly above $W_n$ and $W_m$, touching them only at $x_{n,m}$ and $y_{n,m}$, and strictly above every branch in between. Hence
\begin{equation}
W(\mu)\leq T_{n,m}(\mu)\ \text{on }[L,H],\ \text{with equality only at }x_{n,m},y_{n,m}. \label{eq:global-majorization}
\end{equation}
For any study, $\E[W(\mu_z)]\leq\E[T_{n,m}(\mu_z)]=T_{n,m}(\mu_0)$, because the line is affine and $\E[\mu_z]=\mu_0$. The two-report study attains this bound. Any other study puts positive probability on posteriors where $W<T_{n,m}$ and so falls short. No information gives $W(\mu_0)<T_{n,m}(\mu_0)$. Full revelation gives $(1-p)W(L)+pW(H)<(1-p)T_{n,m}(L)+pT_{n,m}(H)=T_{n,m}(\mu_0)$, since neither endpoint is a contact point.
\end{proof}

\paragraph{Advice versus forecast.} Relative to full revelation, the gain splits into two parts:
\begin{equation}
\underbrace{\E[V(\mu_z^S)]-\E[V(\mu_z^F)]}_{\text{better advice}}
+\underbrace{c\left[\Var(\mu_z^S)-\Var(\mu_z^F)\right]}_{\text{worse forecast}}, \label{eq:welfare-decomposition}
\end{equation}
where $S$ is the optimal study and $F$ full revelation. Full revelation maximizes the variance of posterior means, so when $c>0$ the second part is negative. The interior study wins only because it buys enough better advice to pay for a blurrier forecast.

\begin{remark}[Which equilibrium is played]\label{rem:selection-robustness}
The theorem needs the finest equilibrium only at the two reports the study uses. At every other posterior any partition equilibrium may be played, because no partition equilibrium is worth more than the finest one, and that value already lies below $T_{n,m}$. It therefore holds under the No Incentive to Separate criterion, either player's preferred equilibrium, and ex ante Pareto efficiency at the two reports. It fails if babbling is played at those reports.
\end{remark}

\subsection{The main example}\label{sec:main-example}

\begin{figure}[t]
\centering
\includegraphics[width=0.94\textwidth]{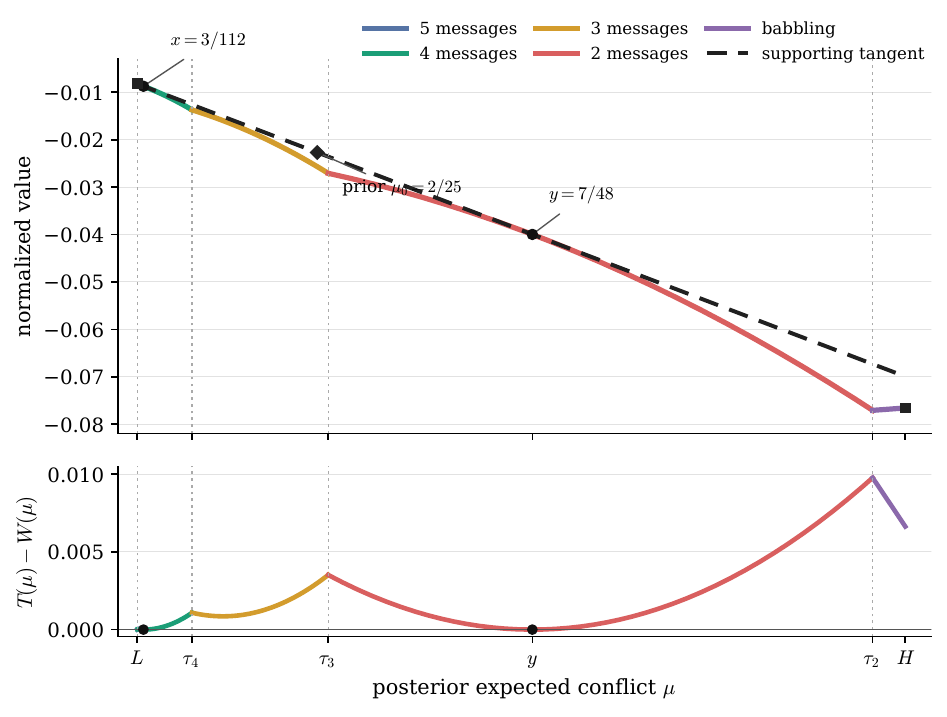}
\caption{\textbf{The main example.} Top: the value $W$ crosses the five-, four-, three-, two-message and babbling regions between $L=249/10000$ and $H=13/50$. The dashed line touches it only at $x=3/112$ (four messages) and $y=7/48$ (two messages). Bottom: the gap between the line and $W$, which is zero only at those two points. Full revelation would use the endpoints, with five messages and no communication.}
\label{fig:global-skipping-example}
\end{figure}

Let
\[
c=\frac1{10},\qquad n=4,\qquad m=2,\qquad L=\frac{249}{10000},\qquad H=\frac{13}{50},\qquad\mu_0=\frac2{25}.
\]
Then $t=13/10$, and the contacts and the line are
\begin{equation}
x=\frac3{112},\qquad y=\frac7{48},\qquad T(\mu)=-\frac{13}{7680}-\frac{21}{80}\mu. \label{eq:example-tangent}
\end{equation}
The support reaches slightly beyond the theorem's window, $L<\tau_5=1/40<x<\tau_4$ and $\tau_3<y<\tau_2=1/4<H$, so full revelation would produce five messages after weak-demand news and babbling after strong-demand news. Direct calculation shows that $T$ still lies above $W$ on all of $[L,H]$, touching it only at $x$ and $y$: on the five-message piece, $T(L)-W_5(L)=1227/500000000>0$ and the gap increases up to $\tau_5$; on the babbling piece the gap is positive and decreasing up to $H$. (Appendix~\ref{app:section4} gives the general version of this extension.) Figure~\ref{fig:global-skipping-example} shows the whole picture.

The optimal study sends the report leading to $x$ with probability $553/1000$ and the report leading to $y$ with probability $447/1000$. The three benchmark values (normalized) are
\[
\text{no information: }-0.02569,\qquad
\text{optimal study: }-0.02269,\qquad
\text{full revelation: }-0.02425.
\]
Relative to full revelation, the optimal study gains about $0.00220$ in advice and loses about $0.00064$ in forecasting, for a net gain of about $0.00156$. Appendix~\ref{app:section4} gives the exact values and an explicit study.

\subsection{What the theorem says}

\emph{Why blur.} After strong-demand news the manager's bias is large and she can say little; after weak-demand news she can say a lot. Revealing demand gives headquarters the best forecast, but it sends the manager into one very informative and one uninformative conversation. Because the value of advice falls in steps, headquarters does better by sending her into two moderately informative conversations. It deliberately leaves demand partly uncertain after both reports.

\emph{Why skip.} Cheap talk hangs just below the convex mediation curve and touches it once per regime (Figure~\ref{fig:protocols}). A straight line joining two distant points near that curve therefore lies above all the scallops in between, so intermediate forecasts are worth less than a mix of two more separated ones. Once one straight line clears several scallops, every regime under it is passed over. The scallops are shallow, which is also why the gains are moderate. In the example, the three-message regime contains the prior, yet the optimal study never sends the manager there.

\subsection{How common, and how large?}\label{sec:how-common}

\begin{figure}[t]
\centering
\includegraphics[width=0.98\textwidth]{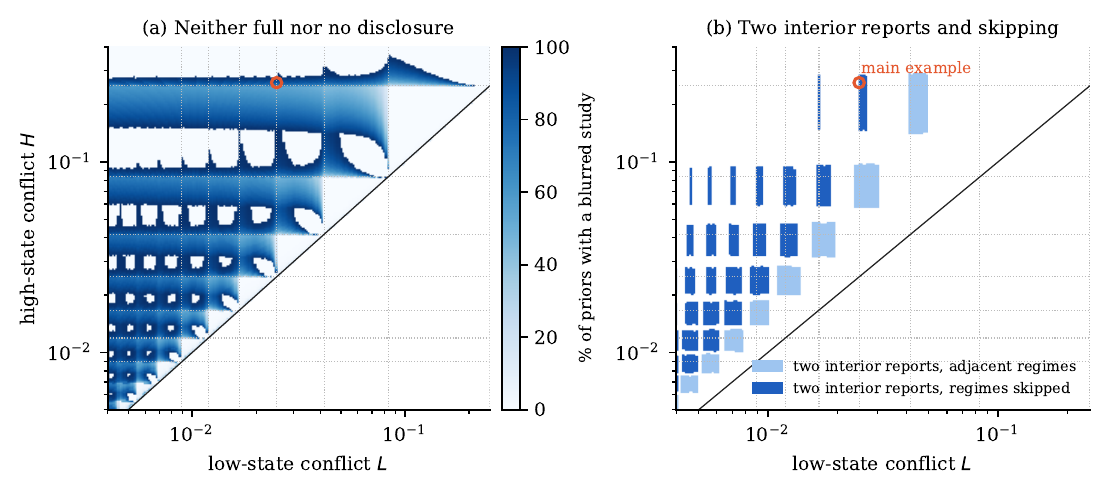}
\caption{\textbf{How common is blurring?} Numerical computation for $c=0.1$ over supports $[L,H]$ (log scales; dotted lines mark the thresholds $\tau_j$). Panel (a): the share of priors in $(L,H)$ for which the optimal study is neither fully revealing nor silent. Panel (b): supports for which, for some prior, the optimal study has two interior reports, in adjacent regimes (light) or skipping at least one regime (dark). The circle marks the main example.}
\label{fig:map}
\end{figure}

Theorem~\ref{thm:global-skipping} gives sufficient conditions, and the main example sits close to the edge of them: the low state can fall by only about $2\times10^{-5}$ before this particular line stops supporting $W$. To see how typical the phenomenon is, we computed the optimal study numerically over a wide grid of supports $[L,H]$, with $0.004\leq L<H\leq0.4$, and over all priors in each support. Figure~\ref{fig:map} shows the result for $c=0.1$.

Three findings stand out.
\begin{itemize}[leftmargin=2.2em]
\item \emph{Blurring is common.} Averaging over the grid and over priors, the optimal study is neither fully revealing nor silent in about $38\%$ of cases ($41\%$ for $c=0$, $32\%$ for $c=0.3$). In most of these cases one report fully reveals one state and the other is interior.
\item \emph{Two interior reports and skipping are less common.} Studies with two interior reports occur in about $7\%$ of cases ($9\%$ for $c=0$, $5\%$ for $c=0.3$), and skipping in about $4\%$ ($6\%$ and $3\%$). They require the low state to sit just inside a regime, which produces the vertical bands in panel (b).
\item \emph{The gains are moderate and largest near babbling.} The best two-interior study improves on the better of full and no disclosure by up to about $10\%$ of the principal's loss ($12\%$ for $c=0$, $5\%$ for $c=0.3$). The largest gains arise when strong demand would push the manager close to babbling ($H$ just below $1/4$): full revelation is then particularly costly, and blurring keeps a two-message conversation alive.
\end{itemize}
The number of skipped regimes can also be made arbitrarily large, but only for conflict states close to zero and with vanishing gains (Appendix~\ref{app:section4}).

\begin{remark}[Further results]\label{rem:section4-extensions}
Appendix~\ref{app:section4} contains (a) a version of Theorem~\ref{thm:global-skipping} whose support extends one regime beyond each contact, as in the example; (b) the extension to non-binary priors, which holds whenever the two-report distribution is a mean-preserving contraction of the prior; and (c) the result that arbitrarily many regimes can be skipped, together with the size of the gain in that case. Appendix~\ref{app:geometric-robustness} shows that the result survives small smooth changes in the expert's type distribution.
\end{remark}

\section{Keeping the Raw Data Private}\label{sec:retained-information}

So far the public report is headquarters' only demand information, so blurring it also blurs headquarters' forecast. In practice, headquarters often keeps the raw market data and releases only a coarser category to the manager. We now allow this.

\subsection{Setup}

The principal privately observes a raw signal $r\sim\kappa(r\mid B)$. Before seeing $r$, she commits to a disclosure rule $\sigma(z\mid r)$; the report $z$ is public. The expert observes $(\theta,z)$ and sends $m$; the principal acts on $(m,z,r)$. Figure~\ref{fig:timing}(b) shows the timing. Write $\alpha\equiv w_P/\delta$, so $c=\alpha^2$, and let $\nu_r\equiv\E[B\mid r]$ be the principal's private forecast and $\mu_z\equiv\E[B\mid z]$ the public one.

\begin{lemma}[Advice depends only on what is public]\label{lem:retained-bias}
The principal's action is $a^*=q(m,z)+\alpha\nu_r$, with $q(m,z)=\E[\theta\mid m,z]$. Given $(\theta,z)$, the part of the expert's expected payoff that depends on her message is $-(q(m,z)-\theta-\mu_z)^2$. The communication game after $z$ is therefore exactly the one of Section~\ref{sec:communication}, with bias equal to the \emph{public} expected conflict $\mu_z$.
\end{lemma}

\begin{proof}
The expert's message depends only on $(\theta,z)$, and $\theta$ is independent of $(B,r,z)$, so $\E[\theta\mid m,z,r]=q(m,z)$. The principal's ideal is $\theta+\alpha B$, so she sets $a^*=q+\alpha\nu_r$. The expert's ideal is $\theta+(\alpha+1)B$, so her error is $q-\theta+\alpha\nu_r-(\alpha+1)B$. Since $\E[\nu_r\mid z]=\mu_z$, the last two terms have conditional mean $-\mu_z$ given $z$. The squared error therefore equals $(q-\theta-\mu_z)^2$ plus a variance term that does not depend on the message.
\end{proof}

The manager does not care what headquarters knows privately; she cares only about what is common knowledge, because that determines the conflict she expects.

\subsection{Acquire everything, disclose selectively}

\begin{proposition}[Acquisition--disclosure separation]\label{prop:separation}
Under the selected equilibrium, the principal's payoff is
\begin{equation}
\E[V(\mu_z)]-c\,\E[\Var(B\mid r)]. \label{eq:retained-objective}
\end{equation}
The first term depends only on the disclosure rule; the second only on the raw data. If raw data are free, the principal optimally acquires full information about $B$ and chooses the public report to maximize $\E[V(\mu_z)]$ over all distributions of posterior means with mean $\mu_0$:
\begin{equation}
\max_{\E[\mu_z]=\mu_0}\E[V(\mu_z)]=\operatorname{cav}_{[L,H]}V(\mu_0). \label{eq:separated-disclosure-problem}
\end{equation}
If $c>0$, every less informative acquisition is strictly worse.
\end{proposition}

\begin{proof}
The principal's error is $(q-\theta)+\alpha(\nu_r-B)$; the two parts are uncorrelated given $(m,z,r)$, which gives \eqref{eq:retained-objective}. With full private information, $\E[\Var(B\mid r)]=0$, and every distribution of public posteriors with mean $\mu_0$ can be generated by garbling $r$ (Appendix~\ref{app:retained-details}). Any other acquisition can only shrink the set of feasible public distributions and, when $c>0$, raises the second term.
\end{proof}

With retained data, disclosure serves communication alone. Disclosure now concavifies $V$ rather than $W=V+c\mu^2$, which is the case $t=1$ of Section~\ref{sec:positive-L}. Let $x^R_{n,m}$, $y^R_{n,m}$ and $T^R_{n,m}$ be \eqref{eq:xnm}--\eqref{eq:Tnm} evaluated at $t=1$:
\begin{equation}
x^R_{n,m}\equiv\frac1{2mn}\sqrt{\frac{m^2-1}{n^2-1}}, \label{eq:retained-x}
\end{equation}
\begin{equation}
y^R_{n,m}\equiv\frac1{2mn}\sqrt{\frac{n^2-1}{m^2-1}}. \label{eq:retained-y}
\end{equation}

\begin{corollary}[Blurring survives private retention]\label{cor:retained-skipping}
Suppose the principal observes $B$ privately. If \eqref{eq:lower-support-condition}--\eqref{eq:upper-support-condition} hold with $x^R_{n,m}$ and $y^R_{n,m}$ in place of $x_{n,m}$ and $y_{n,m}$, then for every $\mu_0\in(x^R_{n,m},y^R_{n,m})$ and every $c\geq0$, the unique optimal public report has two posteriors, $x^R_{n,m}$ and $y^R_{n,m}$, with probabilities as in \eqref{eq:lambda-x}--\eqref{eq:lambda-y}. Arbitrarily many regimes can be skipped.
\end{corollary}

\begin{proof}
Apply Theorem~\ref{thm:global-skipping} and Appendix~\ref{app:section4} with $t=1$, using Proposition~\ref{prop:separation}.
\end{proof}

\subsection{Same primitives, different policy}

Private retention can change the optimal policy, not just its value. Under a common study, a large $c$ makes the forecast so valuable that full revelation is optimal. Under retention, the disclosure problem does not depend on $c$ at all.

\begin{proposition}[Same primitives, different policy]\label{prop:policy-difference}
For every support $0<L<H$ there is a finite threshold $c_F(L,H)$ such that, for every $c>c_F(L,H)$ and every prior, full revelation is the unique optimal common study. Under retained data, the optimal public report is the same for every $c$. Hence, whenever the retained-data optimum is interior, the two institutions choose different public reports for all $c>c_F(L,H)$.
\end{proposition}

\begin{proof}
See Proposition~\ref{prop:full-threshold} in Appendix~\ref{app:retained-details} for the threshold, and Proposition~\ref{prop:separation} for the independence of $c$.
\end{proof}

\begin{figure}[t]
\centering
\includegraphics[width=0.98\textwidth]{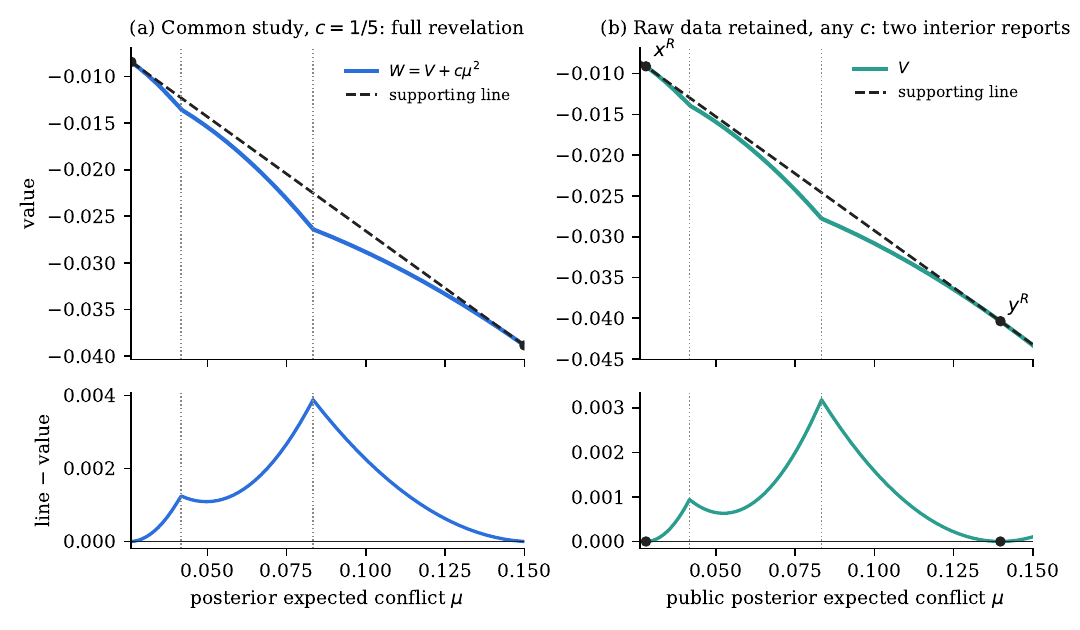}
\caption{\textbf{Same primitives, different policy} ($L=13/500$, $H=3/20$). Panel (a): with a common study and $c=1/5$, the straight line through the endpoints lies above $W$ (bottom: the gap is positive inside), so full revelation is optimal. Panel (b): with retained data, disclosure maximizes the value of advice $V$ alone; the supporting line touches $V$ only at the interior points $x^R$ and $y^R$, whatever the value of $c$.}
\label{fig:retained}
\end{figure}

Take $L=13/500$, $H=3/20$ and $\mu_0=2/25$. Here $c_F=2505/15376\approx0.163$. At $c=1/5$, the common study reveals demand fully. With retained data, the optimal public report has the two interior posteriors $x^R=\sqrt5/80$ and $y^R=\sqrt5/16$, which induce four and two messages, for every $c$. Figure~\ref{fig:retained} shows the two problems.

\begin{table}[t]
\centering
\caption{Two information institutions}
\label{tab:architectures}
\small
\renewcommand{\arraystretch}{1.2}
\begin{tabularx}{\textwidth}{@{}>{\raggedright\arraybackslash}p{0.27\textwidth}YY@{}}
\toprule
 & Common study & Raw data retained \\
\midrule
Demand information at the action & Public report only & Raw data and public report \\
What the public report must balance & Advice and forecast: $\E[V(\mu_z)+c\mu_z^2]$ & Advice only: $\E[V(\mu_z)]$ \\
Optimal report when $c$ is large & Full revelation & Unchanged; can remain interior \\
\bottomrule
\end{tabularx}
\end{table}

\paragraph{Implications.} Acquiring a forecast and making it common knowledge are different decisions. A firm that keeps its raw data can publish a coarse internal forecast without sacrificing its own investment decision; the coarseness is a tool for managing what its managers believe, not a loss of knowledge. This requires the firm to commit to the publication rule before seeing the data, for example through a fixed internal reporting format. Table~\ref{tab:architectures} summarizes the comparison.

\section{When Weak Demand Removes the Conflict}\label{sec:zero-baseline}

Suppose now that the manager and headquarters agree when demand is weak: $B\in\{0,H\}$, so the low state carries no conflict. We maintain the common study and $0\leq c<1$, so that every informative branch is concave.

\paragraph{The key observation.} At zero expected conflict the expert can reveal her information fully, and advice is perfect: $W(0)=0$. Every line used in Section~\ref{sec:positive-L} lies below this point, since $T_{n,m}(0)=-t/(12m^2n^2)<0$. Such a line cannot support $W$ on a domain that contains zero. A report that certifies weak demand is so valuable that it changes the shape of the optimal study.

\begin{proposition}[The zero-conflict benchmark]\label{prop:zero-summary}
Let $B\in\{0,H\}$, $0\leq c<1$ and $\mu_0\in(0,H)$, and exclude the transition values of $H$ listed in Appendix~\ref{app:zero-details}. The unique optimal study takes one of four forms:
\begin{enumerate}[label=\textup{(\alph*)},leftmargin=2.2em]
\item \emph{Certify alignment:} two reports, one revealing that the conflict is zero and one leading to an interior posterior;
\item \emph{No information};
\item \emph{Pool toward high conflict:} two reports, one revealing high conflict $H$ and one leading to an interior posterior;
\item \emph{Full revelation.}
\end{enumerate}
In every case at most one report leads to an interior posterior, so no optimal study places the expert in two distinct interior regimes. When $H\leq1/4$, the only regimes that can be used are the one containing $H$ and the next finer one.
\end{proposition}

\begin{proof}
This summarizes Theorem~\ref{thm:zero-baseline} in Appendix~\ref{app:zero-details}, which gives the closed-form envelope and the optimal study for every $(H,c,\mu_0)$.
\end{proof}

\begin{figure}[t]
\centering
\includegraphics[width=0.98\textwidth]{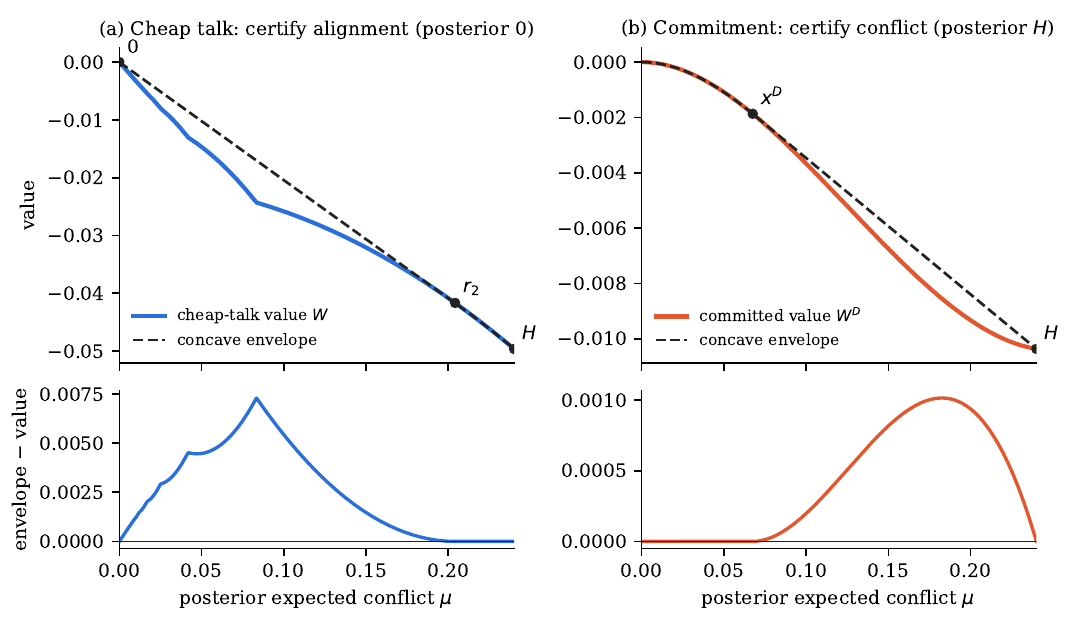}
\caption{\textbf{What the study should certify} ($L=0$, $H=0.24$, $c=1/2$). Panel (a), cheap talk: the envelope runs from the origin to $r_2$ and then follows $W$. The optimal study certifies alignment (posterior $0$, perfect advice) or sends the expert to $r_2$. Panel (b), committed action rule (Section~\ref{sec:delegation}): the value is smooth and concave at low conflict, and the envelope is a line from $x^D$ to $H$. The optimal study now certifies conflict (posterior $H$) or reveals nothing. Bottom panels: the gap between envelope and value.}
\label{fig:zero-baseline}
\end{figure}

Figure~\ref{fig:zero-baseline}(a) shows a typical case. The envelope of $W$ starts at the origin, follows a straight line to a tangency point $r_2$ in the two-message regime, and then follows $W$. For priors below $r_2$, the optimal study sometimes certifies that demand is weak, in which case the manager tells headquarters everything, and otherwise sends her to $r_2$.

\paragraph{Implications.} When some demand news would remove the conflict altogether, the most valuable thing a study can say is ``demand is weak'': that report realigns the manager's interests with headquarters' and restores perfect advice. Regime skipping requires the conflict to be bounded away from zero in both states. Panel (b) of Figure~\ref{fig:zero-baseline} previews Section~\ref{sec:delegation}. With commitment, a cap already handles small conflicts well, so a report certifying alignment is used only as part of full revelation; the distinctive report is the one that certifies \emph{high} conflict.

\section{Committing to the Action Rule}\label{sec:delegation}

So far the principal commits to the public study but not to how she will use the expert's advice. We now let her commit to both at the start. She announces a study together with a rule that fixes her action for every report of the expert and every outcome of the study. This is the delegation problem of \citet{Holmstrom1977,Holmstrom1984} with a public study placed in front of it. The section is a test of our mechanism: once the principal has a direct instrument for managing the expert's bias, does the motive to blur public information survive? We keep the common-study architecture, the uniform type, quadratic payoffs, and binary conflict $B\in\{L,H\}$ with $0\leq L<H$. There are no transfers, and the action rule is deterministic.

\subsection{The problem}

The principal chooses a study $\pi(z\mid s)$ and an action rule $a(\hat\theta,z)$, which gives her action for every report $\hat\theta\in[0,1]$ and every study outcome $z$. The timing is as follows.

\begin{enumerate}[label=\arabic*.,leftmargin=2.2em]
\item Nature draws $(\theta,s)$. The expert observes $\theta$.
\item The principal commits to the study $\pi$ and the rule $a$.
\item The study outcome $z$ is publicly observed.
\item The expert reports $\hat\theta$.
\item The action $a(\hat\theta,z)$ is taken.
\item The state $s$ is revealed and payoffs are realized.
\end{enumerate}

\noindent Figure~\ref{fig:timing}(c) shows this timing. We restrict attention to rules under which the expert reports truthfully.\footnote{This is without loss among deterministic mechanisms in which the expert uses a pure strategy: if $\tilde a(m,z)$ is any rule and $m^*(\theta,z)$ a best response, the rule $a(\theta,z)=\tilde a(m^*(\theta,z),z)$ induces the same outcome and satisfies \eqref{eq:IC-delegation}.} Truth-telling requires, for every outcome $z$ and all types $\theta,\hat\theta$,
\begin{equation}
\E\!\left[u_E(a(\theta,z),\theta,s)\mid\theta,z\right]
\geq
\E\!\left[u_E(a(\hat\theta,z),\theta,s)\mid\theta,z\right]. \label{eq:IC-delegation}
\end{equation}
The principal solves
\begin{equation}
\max_{(\pi,a)}\ \E\!\left[u_P(a(\theta,z),\theta,s)\right]
\quad\text{subject to \eqref{eq:IC-delegation}.} \label{eq:program-P}
\end{equation}

One benchmark organizes the analysis. Suppose the expert's bias is a known number $\mu\geq0$, the type is uniform, and the principal can commit. The best she can achieve is
\begin{equation*}
V^D(\mu)\equiv
\sup_{b:[0,1]\to\mathbb R}
\left\{-\int_0^1\bigl(b(\theta)-\theta\bigr)^2d\theta\ :\
\bigl(b(\theta)-\theta-\mu\bigr)^2\leq\bigl(b(\hat\theta)-\theta-\mu\bigr)^2
\ \ \forall\,\theta,\hat\theta\right\}. \tag{$D_\mu$}\label{eq:delegation-problem}
\end{equation*}
$V^D(\mu)$ is the communication value with commitment. It plays the role that the cheap-talk value $V(\mu)$ played in Section~\ref{sec:communication}.

\subsection{The problem splits posterior by posterior}

\begin{proposition}[Posterior-wise decomposition]\label{prop:delegation-decomposition}
Fix any study $\pi$.
\begin{enumerate}[label=\textup{(\roman*)},leftmargin=2.2em]
\item A rule $a$ satisfies \eqref{eq:IC-delegation} if and only if, after every outcome $z$, the shifted rule $b_z(\theta)\equiv a(\theta,z)-w_Pe_z$ satisfies the constraint of \eqref{eq:delegation-problem} with $\mu=\mu_z$.
\item Every rule satisfying \eqref{eq:IC-delegation} gives the principal at most
\begin{equation}
\E_z\!\left[V^D(\mu_z)-c\Var(B\mid z)\right]. \label{eq:delegation-upper-bound}
\end{equation}
\item The bound is attained by the rule
\begin{equation}
a^*(\theta,z)=
\begin{cases}
\min\{\theta+\mu_z,\,1-\mu_z\}+\sqrt c\,\mu_z, & \mu_z\leq\dfrac12,\\[8pt]
\dfrac12+\sqrt c\,\mu_z, & \mu_z\geq\dfrac12,
\end{cases} \label{eq:optimal-committed-rule}
\end{equation}
which depends on the study outcome only through $\mu_z$.
\end{enumerate}
Hence the principal chooses $\pi$ to maximize \eqref{eq:delegation-upper-bound}. With binary conflict, this is the same as maximizing $\E[W^D(\mu_z)]$ over Bayes-plausible distributions of posterior means on $[L,H]$, where $W^D(\mu)\equiv V^D(\mu)+c\mu^2$; the two objectives differ by the prior-fixed constant $c(L+H)\mu_0-cLH$ of \eqref{eq:W-definition}.
\end{proposition}

\begin{proof}
The study depends only on $s$, so $\theta$ is independent of $(s,z)$. Given $z$, $\theta$ is therefore uniform, and $s$ has mean $e_z$ and variance $\Var(s\mid z)$ whatever $\theta$ is. For any action $x$,
\begin{align*}
\E\!\left[(x-\theta-w_Es)^2\mid\theta,z\right]&=(x-\theta-w_Ee_z)^2+w_E^2\Var(s\mid z),\\
\E\!\left[(x-\theta-w_Ps)^2\mid\theta,z\right]&=(x-\theta-w_Pe_z)^2+w_P^2\Var(s\mid z).
\end{align*}
Write $x=b+w_Pe_z$. Since $\mu_z=(w_E-w_P)e_z$, the expert's first term becomes $(b-\theta-\mu_z)^2$ and the principal's becomes $(b-\theta)^2$. The variance terms do not depend on the report. Part (i) follows.

For (ii), the principal's payoff after $z$ is $-\int_0^1(b_z(\theta)-\theta)^2d\theta-w_P^2\Var(s\mid z)$. By (i) the first term is at most $V^D(\mu_z)$. Also $w_P^2\Var(s\mid z)=c\Var(B\mid z)$, because $B=\delta s$ and $c=(w_P/\delta)^2$. Taking expectations over $z$ gives \eqref{eq:delegation-upper-bound}.

For (iii), Lemma~\ref{lem:delegation-value} below shows that $b^*_\mu(\theta)=\min\{\theta+\mu,1-\mu\}$ for $\mu\leq1/2$, and $b^*_\mu\equiv1/2$ for $\mu\geq1/2$, attains $V^D(\mu)$. Adding back $w_Pe_z=(w_P/\delta)\mu_z=\sqrt c\,\mu_z$ gives \eqref{eq:optimal-committed-rule}, which reaches the bound at every $z$. The binary form follows from \eqref{eq:binary-variance} and $\E[\mu_z]=\mu_0$, exactly as in \eqref{eq:W-definition}.
\end{proof}

Two features matter. First, the expert reports after seeing $z$, so the truth-telling constraints after different outcomes never interact. Committing at the start to the whole rule therefore gives the same value as choosing the best rule after each outcome. Second, that best rule depends on the outcome only through $\mu_z$. At the start, the principal in effect announces the study together with a family of caps indexed by the public posterior. After outcome $z$, the expert may choose any action up to $1-\mu_z+\sqrt c\,\mu_z$ when $\mu_z\leq1/2$, and has no discretion when $\mu_z\geq1/2$.

\subsection{The best committed rule}

\begin{lemma}[Optimal delegation at a known bias]\label{lem:delegation-value}
Every rule satisfying the constraint of \eqref{eq:delegation-problem} is nondecreasing. Moreover:
\begin{itemize}[leftmargin=2.2em]
\item If $\mu\leq1/2$, the best rule is a \emph{cap}: the expert may choose any action up to $1-\mu$. Types $\theta\leq1-2\mu$ receive their ideal $\theta+\mu$, and all higher types receive $1-\mu$.
\item If $\mu\geq1/2$, the best rule ignores the expert and takes the action $1/2$.
\end{itemize}
The resulting value is
\begin{equation}
V^D(\mu)=
\begin{cases}
-\mu^2+\dfrac43\mu^3, & 0\leq\mu\leq\dfrac12,\\[8pt]
-\dfrac1{12}, & \mu\geq\dfrac12.
\end{cases} \label{eq:VD}
\end{equation}
\end{lemma}

The result is classical \citep{Holmstrom1977,MelumadShibano1991,AlonsoMatouschek2008}, and $V^D$ also equals the value of optimal arbitration in \citet{GoltsmanEtAl2009}; Appendix~\ref{app:delegation-proofs} gives a short self-contained proof. The idea is simple. Truth-telling turns any rule into a menu of actions from which each type picks the one closest to her ideal. A type who gets at least her ideal costs the principal at least $\mu^2$, because the expert already leans upward. The only way to do better is to push high types \emph{down}, and the best way to do so is a single ceiling.

\subsection{The optimal public study}

\begin{lemma}[Shape of the design value]\label{lem:WD-shape}
Let $\hat\mu\equiv\max\{(1-c)/4,\,0\}$, which is below $1/2$. The design value
\[
W^D(\mu)=V^D(\mu)+c\mu^2=
\begin{cases}
-(1-c)\mu^2+\dfrac43\mu^3, & 0\leq\mu\leq\dfrac12,\\[8pt]
-\dfrac1{12}+c\mu^2, & \mu\geq\dfrac12,
\end{cases}
\]
is smooth: it is continuously differentiable, with slope $c$ at $\mu=1/2$. It is strictly concave on $[0,\hat\mu]$ and convex on $[\hat\mu,\infty)$. The convexity is strict, except that for $c=0$ the function is constant on $[1/2,\infty)$.
\end{lemma}

\begin{proof}
On $[0,1/2]$, the second derivative $-2(1-c)+8\mu$ is negative below $\hat\mu$ and positive above it. On $[1/2,\infty)$ the slope is $2c\mu$. Both one-sided slopes at $1/2$ equal $c$. Hence the slope is continuous, falls on $[0,\hat\mu]$, and rises on $[\hat\mu,\infty)$ (strictly, except for $c=0$ beyond $1/2$).
\end{proof}

The optimal study is read off the concave envelope of $W^D$ on $[L,H]$. Because $W^D$ is first concave and then convex, this envelope takes one of three forms: $W^D$ itself; the straight line from $(L,W^D(L))$ to $(H,W^D(H))$; or $W^D$ up to some point followed by a straight line to $(H,W^D(H))$. In the third case the switch point is where the tangent to $W^D$ passes through $(H,W^D(H))$. Define the \emph{tangent gap}
\begin{equation}
\varphi(x)\equiv W^D(x)+(W^D)'(x)(H-x)-W^D(H), \label{eq:tangent-gap-H}
\end{equation}
the height of the tangent at $x$, measured at $H$, above $W^D(H)$.

\begin{proposition}[The optimal public study under commitment]\label{prop:delegation-design}
Let $\mu_0\in(L,H)$.\footnote{We exclude $c=0$ with $L\geq1/2$. Then $W^D$ is constant on $[L,H]$ and every study is optimal.} The optimal distribution of posterior means is unique and is as follows.
\begin{enumerate}[label=\textup{(\alph*)},leftmargin=2.2em]
\item \emph{No study.} If $H\leq\hat\mu$, the optimal study reveals nothing.
\item \emph{Full revelation.} If $H>\hat\mu$ and $\varphi(L)\leq0$, the optimal study reveals the state.
\item \emph{Two reports.} If $H>\hat\mu$ and $\varphi(L)>0$, there is a unique $x^D\in(L,\hat\mu)$ with $\varphi(x^D)=0$. If $\mu_0\leq x^D$, the optimal study reveals nothing. If $\mu_0>x^D$, it has two reports, with posteriors $x^D$ and $H$ and probabilities
\begin{equation}
\Prob(\mu_z=H)=\frac{\mu_0-x^D}{H-x^D},
\qquad
\Prob(\mu_z=x^D)=\frac{H-\mu_0}{H-x^D}. \label{eq:delegation-weights}
\end{equation}
\end{enumerate}
When $H\leq1/2$, everything is explicit:
\begin{equation}
\varphi(x)=-(H-x)^2\left[\frac83x+\frac43H-(1-c)\right],
\qquad
x^D=\frac{3(1-c)}8-\frac H2, \label{eq:xD-closed-form}
\end{equation}
and full revelation is optimal exactly when $x^D\leq L$, that is, when $8L+4H\geq3(1-c)$. When $H>1/2$, $x^D$ solves
\begin{equation}
(H-x)^2\left[\frac83x+\frac43H-(1-c)\right]=\frac{(2H-1)^2(4H+1)}{12}. \label{eq:phi-closed-form}
\end{equation}
If $c\geq1$, then $\hat\mu=0$ and full revelation is always optimal.
\end{proposition}

The proof, in Appendix~\ref{app:delegation-proofs}, uses the supporting-line argument of Theorem~\ref{thm:global-skipping}.

\begin{remark}[No interior two-point designs]\label{rem:delegation-no-skipping}
Under commitment, at most one posterior of the optimal study is interior, and when there are two posteriors the other one is the extreme $H$. The two interior posteriors of Theorem~\ref{thm:global-skipping} come from the upward kinks of the cheap-talk value at the communication thresholds. Commitment removes those kinks. The only reason left to spread public beliefs is the single convex region of $W^D$ at high conflict.
\end{remark}

\paragraph{When weak demand removes the conflict.} Proposition~\ref{prop:delegation-design} also covers $L=0$. Because $W^D$ is concave at low conflict, a report revealing zero conflict is used only as part of full revelation. Compare Section~\ref{sec:zero-baseline}: under cheap talk, certifying alignment is valuable because it restores perfect advice; under commitment, small conflicts are already handled well by a cap, and the distinctive report certifies \emph{high} conflict instead. Figure~\ref{fig:zero-baseline} contrasts the two.

\subsection{What Proposition~\ref{prop:delegation-design} says}

\emph{Why the value has this shape.} With commitment, the principal controls the expert's discretion directly through a cap. When conflict is small, the cap binds only for the highest types, and every other type gets her own ideal action. The loss is then about $\mu^2$: a little conflict costs little, and each further unit costs more. When conflict is large, the cap pools almost every type. The loss can never exceed $1/12$, the loss from ignoring the expert, so each further unit of conflict costs less, and nothing once $\mu\geq1/2$. The value of delegation therefore falls at an increasing rate at low conflict and at a decreasing rate at high conflict. The direct-information term $c\mu^2$ adds convexity and moves the switch point $\hat\mu=(1-c)/4$ to the left.

\begin{figure}[t]
\centering
\includegraphics[width=0.98\textwidth]{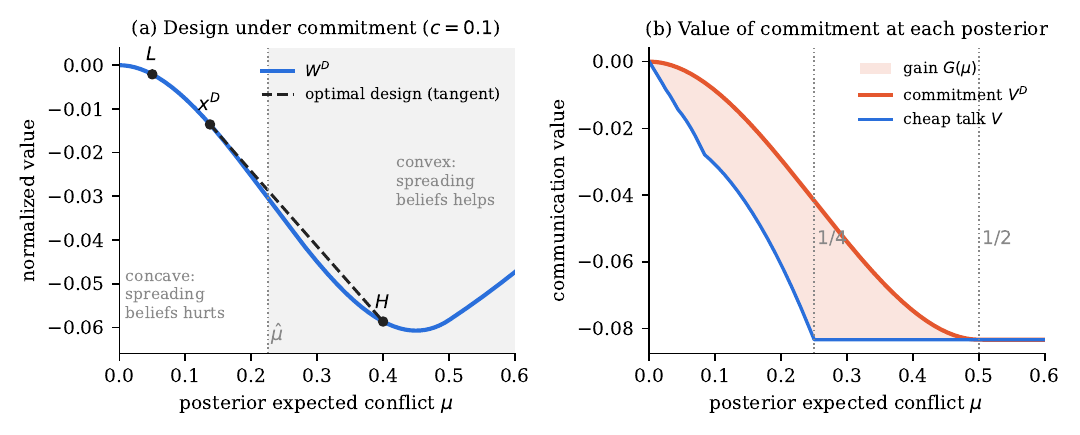}
\caption{\textbf{Public design under commitment.} Panel (a): $W^D$ for $c=0.1$ is concave on $[0,\hat\mu]$ and convex beyond (shaded). With $L=0.05$ and $H=0.4$, the optimal envelope follows $W^D$ up to $x^D=0.1375$ and then the tangent through $(H,W^D(H))$. Priors above $x^D$ are split between $x^D$ and $H$; priors below $x^D$ receive no information. Panel (b): the communication value with commitment, $V^D$, and without it, $V$. The shaded difference $G(\mu)$ is the gain from commitment at each posterior; it is positive exactly when $0<\mu<1/2$.}
\label{fig:delegation-design}
\end{figure}

\begin{figure}[t]
\centering
\includegraphics[width=0.98\textwidth]{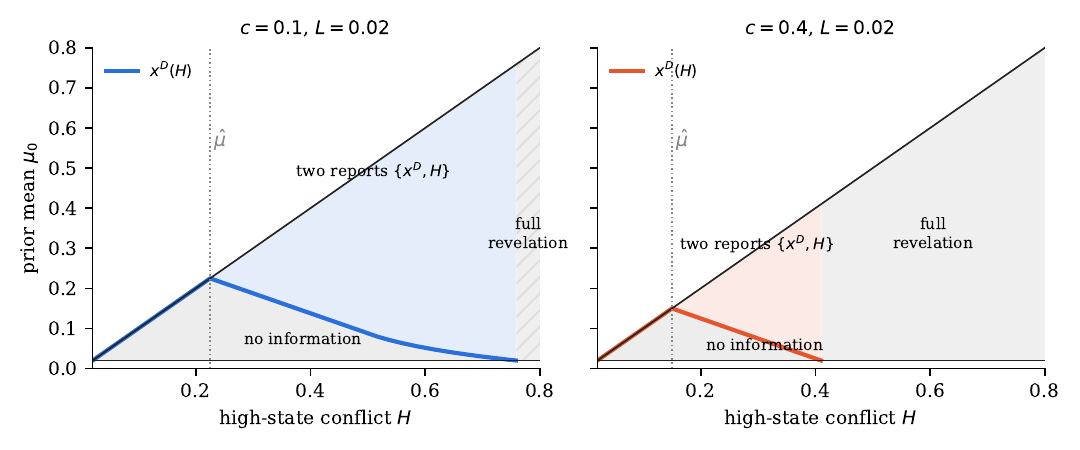}
\caption{\textbf{The optimal public study under commitment} for $L=0.02$. For each high-state conflict $H$ and prior mean $\mu_0\in(L,H)$, the panel shows whether the optimal study reveals nothing (below the curve $x^D(H)$), has two reports $\{x^D,H\}$ (above it), or reveals the state (hatched, where $x^D$ reaches $L$). A larger $c$ moves $\hat\mu$ and $x^D(H)$ down and enlarges the full-revelation region: full revelation starts at $H\approx0.76$ for $c=0.1$ and at $H\approx0.41$ for $c=0.4$.}
\label{fig:delegation-policy}
\end{figure}

\emph{What the principal does.} Where the value is concave, spreading public beliefs mixes a better and a worse delegation outcome, and the mixture is worth less than staying at the prior. Where it is convex, the mixture is worth more. Figure~\ref{fig:delegation-design}(a) shows the resulting envelope. Proposition~\ref{prop:delegation-design} then reads as follows.
\begin{itemize}[leftmargin=2.2em]
\item If even the high state's conflict $H$ is in the concave zone, the principal runs no study.
\item If the two states are far apart, or $c$ is large, she reveals the state.
\item Otherwise the study has two reports. One is issued only in the high state and certifies it, so the posterior is $H$. The other pools the low state with the remaining high states and leaves posterior conflict at $x^D$, inside the concave zone. In the capacity application, a certified strong-demand report gives the manager little or no discretion, while the pooled report gives her a wide cap. If the prior is already below $x^D$, no study is run.
\end{itemize}
When $H\leq1/2$, \eqref{eq:xD-closed-form} shows that $x^D$ falls as $H$ or $c$ rises: the principal discloses more when the high state is more extreme or when demand matters more for her own decision. Figure~\ref{fig:delegation-design}(b) compares the value with and without commitment, and Figure~\ref{fig:delegation-policy} maps the three policies.

\subsection{The gain from commitment}

Let $\Pi^C$ be the principal's best ex ante payoff when she commits only to the study and the finest cheap-talk equilibrium is played, as in Sections~\ref{sec:communication}--\ref{sec:zero-baseline}. Let $\Pi^D$ be her best ex ante payoff when she commits to the study and the action rule. At each posterior, the gain from commitment is
\begin{equation}
G(\mu)\equiv V^D(\mu)-V(\mu),\qquad\mu\geq0. \label{eq:commitment-gain}
\end{equation}
Both design problems have optimal studies, because $W$ and $W^D$ are continuous on $[L,H]$ and the set of Bayes-plausible distributions of posterior means on $[L,H]$ is compact. Denote any optimal distributions of posterior means by $\eta^C$ (without commitment) and $\eta^D$ (with commitment).

\begin{corollary}[Gain from commitment]\label{cor:commitment-gain}
Let $\mu_0\in(L,H)$.
\begin{enumerate}[label=\textup{(\roman*)},leftmargin=2.2em]
\item At each posterior, commitment helps exactly when conflict is strictly between $0$ and $1/2$: $G(\mu)>0$ for $0<\mu<1/2$, and $G(\mu)=0$ for $\mu=0$ and for $\mu\geq1/2$.
\item The total gain lies between the average gains at the two optimal studies:
\begin{equation}
\E_{\eta^C}\bigl[G(\mu_z)\bigr]\ \leq\ \Pi^D-\Pi^C\ \leq\ \E_{\eta^D}\bigl[G(\mu_z)\bigr]. \label{eq:gain-bounds}
\end{equation}
\item Commitment gives no gain if $L\geq1/2$. It gives a strict gain whenever the optimal cheap-talk study puts positive probability on posteriors in $(0,1/2)$, and in particular whenever $L>0$ and $\mu_0<1/2$.
\end{enumerate}
\end{corollary}

\begin{proof}
(i) Both values are zero at $\mu=0$ and equal $-1/12$ for $\mu\geq1/2$. For $\mu$ in the $j$-message region $I_j$ with $j\geq2$, subtracting \eqref{eq:Vj} from \eqref{eq:VD} gives
\[
G(\mu)=\frac1{12j^2}+\frac{j^2-4}{3}\mu^2+\frac43\mu^3>0.
\]
In the babbling region $1/4\leq\mu<1/2$,
\[
G(\mu)=\frac1{12}-\mu^2+\frac43\mu^3=\frac{(1-2\mu)^2(1+4\mu)}{12}>0.
\]
(That $G\geq0$ also follows directly: every cheap-talk partition equilibrium is itself a feasible committed rule.)

(ii) Both payoffs equal the corresponding normalized design value minus the same prior-fixed constant. If the principal runs the cheap-talk optimal study $\eta^C$ but commits to the best rule \eqref{eq:optimal-committed-rule}, she gets $\E_{\eta^C}[W^D]=\E_{\eta^C}[W]+\E_{\eta^C}[G]$; hence $\Pi^D-\Pi^C\geq\E_{\eta^C}[G]$. If she runs $\eta^D$ without commitment, she gets $\E_{\eta^D}[W]=\E_{\eta^D}[W^D]-\E_{\eta^D}[G]$; hence $\Pi^C\geq\Pi^D-\E_{\eta^D}[G]$.

(iii) If $L\geq1/2$, then $G=0$ on $[L,H]$ by (i). If $\eta^C$ puts positive probability on $(0,1/2)$, the lower bound in \eqref{eq:gain-bounds} is positive by (i). If $L>0$ and $\mu_0<1/2$, then because $\eta^C$ has mean $\mu_0<1/2$, it puts positive probability on $[L,1/2)$, which lies inside $(0,1/2)$.
\end{proof}

\emph{The main example under commitment.} Return to the main example of Section~\ref{sec:main-example}: $c=1/10$, $L=249/10000$, $H=13/50$, $\mu_0=2/25$. Here $H\leq1/2$ and $x^D=83/400>L$, so case (c) of Proposition~\ref{prop:delegation-design} applies. Because $\mu_0<x^D$, the optimal study under commitment reveals \emph{nothing}. Its normalized value is $W^D(\mu_0)=-238/46875\approx-0.005077$, against $T(\mu_0)=-4357/192000\approx-0.022693$ without commitment. The gain is
\[
\Pi^D-\Pi^C=\frac{140923}{8000000}\approx0.017615.
\]
It lies between the bounds in \eqref{eq:gain-bounds}, $\E_{\eta^C}[G]=52979/3386880\approx0.015642$ and $G(\mu_0)=34777/1687500\approx0.020609$. In original payoffs, the principal's expected loss falls from about $0.024325$ to about $0.006709$, a reduction of about $72\%$. Commitment also changes what the study is for. Without commitment, the study is valuable because it reshapes cheap talk. With commitment, the cap does that job directly, and the study is not run.

\section{The Protocol Determines the Design}\label{sec:protocols}

We can now put the three protocols side by side. In each, the principal commits to the public study. Under cheap talk she commits to nothing else. Under mediation, the expert speaks to a neutral mediator who recommends an action, but the principal still chooses freely. Under commitment, she fixes her action rule in advance. The value of advice at a known conflict is $V$, $V^M$ and $V^D$ respectively, and they are ranked by how much the principal can commit to:
\begin{equation}
V(\mu)\leq V^M(\mu)\leq V^D(\mu),\qquad V^D(\mu)-V^M(\mu)=\frac{\mu(1-2\mu)^2}{3}\ \text{ for }\mu\leq\tfrac12. \label{eq:protocol-ranking}
\end{equation}
The first inequality is Lemma~\ref{lem:scallops}; the identity follows from \eqref{eq:VM} and \eqref{eq:VD}.

\begin{proposition}[Mediation: reveal the conflict]\label{prop:mediation}
Suppose that after every report the expert communicates through an optimal mediator. Then the principal's design value $W^M=V^M+c\mu^2$ is convex, and fully revealing the conflict is optimal for every prior. It is uniquely optimal unless $c=0$ and $L\geq1/2$. The same holds with retained data, where the design value is $V^M$.
\end{proposition}

\begin{proof}
On $[0,1/2]$, $W^M(\mu)=-\frac\mu3+\frac{1+3c}3\mu^2$ is strictly convex; on $[1/2,\infty)$, $W^M=-\frac1{12}+c\mu^2$ is convex; both slopes equal $c$ at $1/2$. So $W^M$ is convex, and the straight line through $(L,W^M(L))$ and $(H,W^M(H))$ lies above it. It touches $W^M$ inside $(L,H)$ only if $W^M$ is linear on $[L,H]$, which happens only when $c=0$ and $L\geq1/2$. Full revelation therefore attains the line, uniquely in all other cases. With retained data, set $c=0$ in this argument.
\end{proof}

Table~\ref{tab:protocols} collects the three answers.

\begin{table}[t]
\centering
\caption{The protocol determines the optimal public study}
\label{tab:protocols}
\small
\renewcommand{\arraystretch}{1.25}
\begin{tabularx}{\textwidth}{@{}>{\raggedright\arraybackslash}p{0.15\textwidth}>{\raggedright\arraybackslash}p{0.2\textwidth}YY@{}}
\toprule
Protocol & Shape of the value of advice in conflict & Optimal public study & Most valuable news \\
\midrule
Mediation & Convex & Reveal the conflict fully (Prop.~\ref{prop:mediation}) & All news about the conflict \\
Cheap talk & Convex trend minus one scallop per regime & Blur: never at a threshold, informative between touch points, possibly skipping regimes (Section~\ref{sec:positive-L}) & News that removes the conflict (Section~\ref{sec:zero-baseline}) \\
Commitment & Concave, then convex & Reveal nothing at low conflict; certify high conflict (Prop.~\ref{prop:delegation-design}) & News that the conflict is large \\
\bottomrule
\end{tabularx}
\end{table}

\paragraph{Why the shapes differ.} The answer turns on how costly the \emph{first} units of conflict are. Under mediation they are costly: the loss $\mu(1-\mu)/3$ rises steeply and then levels off. A lottery over conflict therefore lowers the expected loss, and the principal wants as much spread as possible. Under commitment they are cheap: a cap absorbs small conflicts, and the loss rises like $\mu^2$. Spreading small conflicts raises the expected loss, so the principal keeps quiet unless the conflict may be large enough to reach the convex region. Cheap talk sits in between. Its trend is that of mediation, but within each regime it loses a scallop, and the scallops are concave. Locally the principal dislikes spreading; across regimes she likes it. The result is blurring.

\paragraph{More commitment does not mean more information.} By \eqref{eq:protocol-ranking}, the value of advice rises with the principal's commitment power. The value of public information about the conflict does not. It is positive under mediation whenever the conflict is uncertain, positive but selective under cheap talk, and often zero under commitment, as in our main example. Public information about the conflict and commitment over the use of advice are partial substitutes. With a neutral intermediary, by contrast, public information about the conflict is always valuable.

\section{Discussion and Conclusion}\label{sec:conclusion}

\subsection{Three institutions compared}

Table~\ref{tab:institutions} compares the institutions in the example of Figure~\ref{fig:overview}, using the principal's original expected payoff. The common study gains little over full revelation here, because on this support full revelation already uses four and two messages; the main example of Section~\ref{sec:main-example} shows a larger gain. Keeping the raw data adds a further small improvement. Commitment does far better, cutting the expected loss by about three quarters, and makes the study worthless.

\begin{table}[t]
\centering
\caption{The example of Figure~\ref{fig:overview}: $c=1/10$, $L=13/500$, $H=3/20$, $\mu_0=2/25$}
\label{tab:institutions}
\small
\renewcommand{\arraystretch}{1.2}
\begin{tabularx}{\textwidth}{@{}Ylr@{}}
\toprule
Institution and study & Public posteriors & Expected payoff \\
\midrule
Cheap talk, no study & $\mu_0$ (3 messages) & $-0.02670$ \\
Cheap talk, common study, full revelation & $L,H$ (4 and 2 messages) & $-0.02372$ \\
Cheap talk, common study, optimal & $3/112,\ 7/48$ (4 and 2 messages) & $-0.02371$ \\
Cheap talk, raw data retained, optimal disclosure & $\sqrt5/80,\ \sqrt5/16$ (4 and 2 messages) & $-0.02366$ \\
Committed action rule, optimal study & $\mu_0$ (no study) & $-0.00610$ \\
\bottomrule
\end{tabularx}
\end{table}

\subsection{Implications}

\emph{The protocol decides.} Whether and how to inform depends on how the subsequent conversation is organized: reveal under mediation, blur under cheap talk, stay quiet unless conflict is large under commitment (Section~\ref{sec:protocols}).

\emph{Public forecasts crowd out advice.} A forecast that makes the principal more confident about demand can also widen her disagreement with the adviser who knows local conditions. A study must therefore be judged together with the conversation that follows it. When the value of advice falls in steps, the best study can blur the forecast on purpose and place the adviser in two quite different conversations rather than in one middling one (Theorem~\ref{thm:global-skipping}).

\emph{Data governance.} The cost of blurring is a feature of the common-study institution. A firm that keeps its raw data and commits to a coarse publication rule manages its managers' beliefs at no cost to its own decisions (Proposition~\ref{prop:separation}). Acquiring a forecast and making it common knowledge are separate choices, and they can call for opposite policies under the same primitives (Proposition~\ref{prop:policy-difference}).

\emph{Commitment and information design are substitutes.} When the principal can commit to how she uses advice, a cap on discretion handles the bias directly. The value of advice becomes smooth, the regime-skipping motive disappears, and in both of our leading examples no study is run (Section~\ref{sec:delegation}). In this model, information design matters most where commitment is unavailable.

\emph{What to certify.} Under cheap talk, the most valuable news is news that removes the conflict, because it restores full communication (Section~\ref{sec:zero-baseline}). Under commitment, the most valuable news is news that the conflict is large, because it tells the principal when to take discretion away (Proposition~\ref{prop:delegation-design}).

\subsection{Fiscal councils, spending agencies and fiscal rules}\label{sec:fiscal-implications}

Read through the application of Section~\ref{sec:political-application}, the results give five implications for fiscal governance.

\emph{Publishing the official forecast in broad categories can help budgeting.} A precise, optimistic official forecast widens the gap between what a spending agency wants and what the treasury wants, and makes the agency's account of implementation conditions less credible. A forecast published as broad scenarios (for example, ``strong'' versus ``moderate'' growth), rather than as a precise point estimate, can keep the agency's advice informative. Its purpose is not to hide information from the public but to keep the budget conversation useful. Proposition~\ref{prop:no-threshold} adds that the official categories should never leave agencies at the margin where their advice is about to become less informative.

\emph{Treasuries can know more than they publish.} A treasury that keeps its own detailed forecast can let the fiscal council publish coarse categories at no cost to its own budget decisions (Proposition~\ref{prop:separation}). This requires committing to the publication format before the data arrive, which a statutory mandate for the fiscal council can provide.

\emph{Intermediaries make precise forecasts valuable.} If agencies' bids pass through an independent body that acts like a mediator, receiving the agency's information and passing on a recommendation, the model favours publishing the official forecast as precisely as possible (Proposition~\ref{prop:mediation}).

\emph{Fiscal rules and coarse forecasts are substitutes.} A fiscal rule that caps an agency's budget as a function of the official forecast is the committed rule of Section~\ref{sec:delegation}. Under such a rule the forecast loses most of its strategic role, so the case for coarse official forecasts is strongest where fiscal rules are weak or absent.

\emph{What the official forecast should certify depends on the regime.} Under discretionary budgeting, the most valuable official news is news that aligns the agency's interests with the treasury's, for example a downturn in which both agree on a small programme. Under a fiscal rule, the most valuable news is news that the conflict is large, for example a boom in which the cap should bind.

The same logic applies when a legislature consults a specialized committee while an independent budget office publishes economic projections: projections that sharpen the committee's disagreement with the floor can reduce what the floor learns from it, and restrictive procedures that commit the floor play a role similar to the cap.

\subsection{Limitations}

Our results use the uniform--quadratic model, a binary conflict state for the main characterizations, and the finest cheap-talk equilibrium at the reports the study uses; they fail if babbling is played there. Blurring is common in our numerical map, but two interior reports and regime skipping arise in only about $7\%$ and $4\%$ of cases, and their gains are moderate, at up to about $10\%$ of the principal's loss. Appendix~\ref{app:geometric-robustness} shows robustness to small smooth changes in the type distribution, and Appendix~\ref{app:section4} extends the result to non-binary priors under a convex-order condition. We also keep demand independent of local conditions. If the two are correlated, the public report also changes the principal's beliefs about local conditions and the distribution of the expert's type. When the correlation is weak, each report leaves the type distribution close to uniform, and we expect the perturbation argument of Appendix~\ref{app:geometric-robustness} to carry over; a formal treatment, and the case of strong correlation, are left for future work. The commitment analysis covers the common-study institution only; combining commitment with privately retained data, which makes the expert face a lottery over actions, is a natural next step.

\appendix

\section{Details for Section~\ref{sec:positive-L}}\label{app:section4}

\subsection{Algebra for Lemma~\ref{lem:common-tangent}}\label{app:tangent-algebra}

Define
\[
d_{n,m}\equiv
\frac{\sqrt{(n^2-t)(m^2-t)}}{3mn}.
\]
Using \eqref{eq:xnm} and \eqref{eq:ynm},
\[
W_n'(x_{n,m})
=-\frac{2(n^2-t)}3x_{n,m}
=-d_{n,m}
=-\frac{2(m^2-t)}3y_{n,m}
=W_m'(y_{n,m}).
\]
Moreover,
\[
W_n(x_{n,m})+d_{n,m}x_{n,m}
=-\frac{t}{12m^2n^2}
=W_m(y_{n,m})+d_{n,m}y_{n,m},
\]
which proves that \eqref{eq:Tnm} is the common tangent.

Squaring $x_{n,m}>\tau_{n+1}$ and simplifying gives
\[
m^2(2n+1)>t\bigl((n+1)^2-m^2\bigr),
\]
equivalent to \eqref{eq:x-lower-condition}. Likewise, $y_{n,m}<\tau_m$ is equivalent to
\[
n^2(2m-1)>t\bigl(n^2-(m-1)^2\bigr),
\]
which gives \eqref{eq:y-upper-condition}. After squaring, the remaining two regime inequalities reduce to
\[
m^2(2n-1)+t\bigl((n-1)^2-m^2\bigr)>0
\]
and
\[
n^2(2m+1)+t\bigl(n^2-(m+1)^2\bigr)>0.
\]
Both hold under $n>m$ and $t<m^2$.

For $m<k<n$, the gap from the common tangent is
\begin{align*}
T_{n,m}(\mu)-W_k(\mu)
={}&
\frac{k^2-t}{3}\mu^2-d_{n,m}\mu
+\frac1{12k^2}-\frac{t}{12m^2n^2}.
\end{align*}
Its unique minimizer is
\[
\mu=
\frac{\sqrt{(n^2-t)(m^2-t)}}
{2mn(k^2-t)}.
\]
Substitution yields
\[
\frac{
t(k^2-m^2)(n^2-k^2)
}{
12k^2m^2n^2(k^2-t)
},
\]
which is strictly positive.

The two square identities in \eqref{eq:n-contact-square} hold because each contact branch is a quadratic with leading coefficient $-(k^2-t)/3$, $k\in\{n,m\}$, and $T_{n,m}$ is tangent to it at the stated point.

\subsection{Support beyond the contact regimes, general priors, and implementation}\label{app:adjacent}

\begin{proposition}[Support across adjacent thresholds]\label{prop:adjacent-extension}
Suppose $t<m^2$ and the two contacts lie strictly in their selected regimes:
\[
\tau_{n+1}<x_{n,m}<\tau_n,
\qquad
\tau_{m+1}<y_{n,m}<\tau_m.
\]
There are $\varepsilon_->0$ and $\varepsilon_+>0$ such that $T_{n,m}$ dominates $W$ on
\begin{equation}
[\tau_{n+1}-\varepsilon_-,\,\tau_m+\varepsilon_+], \label{eq:extended-support-window}
\end{equation}
with equality only at $x_{n,m},y_{n,m}$. The lower extension can be chosen within the $(n+1)$-message regime and the upper extension within the $(m-1)$-message regime, with $m-1=1$ denoting babbling. Thus for any $L$ just below $\tau_{n+1}$, $H$ just above $\tau_m$, and $\mu_0\in(x_{n,m},y_{n,m})$, the two-contact experiment remains uniquely optimal. Full revelation instead induces $n+1$ messages at $L$ and $m-1$ messages at $H$.
\end{proposition}

\begin{proof}
At the two thresholds, continuity of the selected value and the contact-square identities give
\begin{align*}
T_{n,m}(\tau_{n+1})-W(\tau_{n+1})
&=\frac{n^2-t}{3}(x_{n,m}-\tau_{n+1})^2>0,\\
T_{n,m}(\tau_m)-W(\tau_m)
&=\frac{m^2-t}{3}(\tau_m-y_{n,m})^2>0.
\end{align*}
The first strict gap persists on a sufficiently short interval to the left of $\tau_{n+1}$; the second persists on one to the right of $\tau_m$. Choose both intervals within the neighboring selected regimes and keep the lower endpoint positive. On $[\tau_{n+1},\tau_m]$, the contact-square identities and Lemma~\ref{lem:intermediate-domination} already give domination with equality only at the two contacts. The affine-majorant and equality-set argument in Theorem~\ref{thm:global-skipping} now applies on the larger support.
\end{proof}

The binary prior is particularly convenient because any Bayes-plausible two-contact law is implementable. For a more general prior, implementability imposes one additional condition.

\begin{corollary}[Conditional extension to general priors]\label{cor:general-prior}
Let $B$ have an arbitrary prior law $F_B$ supported on $[L,H]$, with mean $\mu_0\in(x_{n,m},y_{n,m})$. Suppose $T_{n,m}\geq W$ throughout $[L,H]$, with equality only at $x_{n,m},y_{n,m}$. Define
\[
\eta^*=\lambda_x\delta_{x_{n,m}}+\lambda_y\delta_{y_{n,m}},
\qquad
\lambda_x=\frac{y_{n,m}-\mu_0}{y_{n,m}-x_{n,m}},
\quad
\lambda_y=\frac{\mu_0-x_{n,m}}{y_{n,m}-x_{n,m}}.
\]
The two-contact law is implementable if and only if $\eta^*\preceq_{\mathrm{cx}}F_B$: it is a mean-preserving contraction of the prior. Whenever this holds, it is the unique optimal posterior-mean law.
\end{corollary}

\begin{proof}
Every posterior-mean law is dominated by the prior in convex order. Conversely, the martingale-coupling characterization of convex order implements $\eta^*$ as a signal with $\E[B\mid\mu_z]=\mu_z$. Any feasible law satisfies $\E[W(\mu_z)]\leq\E[T_{n,m}(\mu_z)]=T_{n,m}(\mu_0)$. The implementable two-contact law attains this bound. Equality requires support only at the two contacts, and its mean fixes the weights.
\end{proof}

In one dimension, the extra condition is equivalent to
\begin{equation}
\lambda_x(x_{n,m}-q)_+ +\lambda_y(y_{n,m}-q)_+
\leq\E[(B-q)_+]
\quad\text{for every }q\in\mathbb R. \label{eq:general-prior-call}
\end{equation}
The binary endpoint prior automatically satisfies it; a less dispersed continuous prior need not. This is a conditional extension of the optimal law, not a claim that every prior implements it.

To implement the optimal posterior distribution, write
\[
\lambda_y\equiv
\frac{\mu_0-x_{n,m}}{y_{n,m}-x_{n,m}},
\qquad
q_x\equiv\frac{x_{n,m}-L}{H-L},
\qquad
q_y\equiv\frac{y_{n,m}-L}{H-L}.
\]
Specializing \eqref{eq:binary-implementation} gives
\begin{equation}
\Prob(y\mid H)=\frac{\lambda_yq_y}{p},
\qquad
\Prob(y\mid L)=\frac{\lambda_y(1-q_y)}{1-p}, \label{eq:two-contact-implementation}
\end{equation}
with complementary probabilities assigned to signal $x$.

\subsection{Arbitrarily many skipped regimes}\label{app:arbitrary}

\begin{corollary}[Arbitrarily many skipped regimes]\label{cor:arbitrary-skipping}
For every finite $c\geq0$ and every integer $K\geq1$, there exist integers $n>m\geq2$, a binary support $0<L<H$, and an open interval of prior means for which the unique optimal distribution of posterior means skips at least $K$ finite communication regimes.
\end{corollary}

\begin{proof}
Fix $t=1+3c$ and choose an integer
\begin{equation}
m>\max\{t,3,K\}. \label{eq:m-construction}
\end{equation}
Set $n=2m$. The two nonautomatic contact bounds in Lemma~\ref{lem:common-tangent} become
\begin{align}
\frac{m^2(4m+1)}{3m^2+4m+1},
\qquad
\frac{4m^2(2m-1)}{3m^2+2m-1}. \label{eq:2m-bounds}
\end{align}
For $m\geq4$, the first exceeds $m$ because
\[
m^2(4m+1)-m(3m^2+4m+1)
=m(m^2-3m-1)>0.
\]
The second exceeds $m$ because
\[
4m^2(2m-1)-m(3m^2+2m-1)
=m(5m-1)(m-1)>0.
\]
Both expressions in \eqref{eq:2m-bounds} therefore exceed $m>t$, and $t<m^2$ also holds.

Choose, for example,
\begin{equation}
L=\frac{\tau_{2m+1}+x_{2m,m}}2,
\qquad
H=\frac{y_{2m,m}+\tau_m}2, \label{eq:support-construction}
\end{equation}
and take
\[
\mu_0=\frac{x_{2m,m}+y_{2m,m}}2.
\]
All inequalities in Theorem~\ref{thm:global-skipping} are strict. The optimal experiment moves from regime $2m$ to regime $m$ and skips
\[
2m-m-1=m-1\geq K
\]
intermediate regimes.

For fixed $t$,
\begin{align}
x_{2m,m}
&=
\frac1{4m^2}
\sqrt{\frac{m^2-t}{4m^2-t}}
\sim\frac1{8m^2}, \label{eq:x-asymptotic}\\
y_{2m,m}
&=
\frac1{4m^2}
\sqrt{\frac{4m^2-t}{m^2-t}}
\sim\frac1{2m^2}. \label{eq:y-asymptotic}
\end{align}
Thus the constructed supports approach zero as the number of skipped regimes grows. More generally, this convergence is forced along any sequence satisfying the theorem for fixed $c$ and skipping an unbounded number of regimes. Such a sequence has $n\to\infty$. If $m$ remained bounded, the right side of \eqref{eq:x-lower-condition} would converge to zero, contradicting the fixed inequality $t\geq1$. Hence $m\to\infty$, and the support condition $0<L<H<\tau_m$ then gives $L,H\to0$. Finally, strict inequalities and continuity of the contacts imply that the result persists on an open set of $(c,L,H,\mu_0)$, relative to the parameter space $c\geq0$.
\end{proof}

The existence construction is quantitatively informative only with its scale attached. The number of bypassed regimes grows, but the conflict states and the payoff difference shrink.

\begin{proposition}[Scale of extensive skipping]\label{prop:skipping-scale}
Fix finite $t=1+3c$, and take $n=2m$ with the support and midpoint prior in \eqref{eq:support-construction}. As $m\to\infty$,
\begin{align}
x_{2m,m}&=\frac1{8m^2}-\frac{3t}{64m^4}+O(m^{-6}),
&y_{2m,m}&=\frac1{2m^2}+\frac{3t}{16m^4}+O(m^{-6}), \label{eq:contact-scale}\\
L_m&=\frac1{8m^2}-\frac1{32m^3}+\frac{2-3t}{128m^4}+O(m^{-5}),
&H_m&=\frac1{2m^2}+\frac1{4m^3}+\frac{8+3t}{32m^4}+O(m^{-5}). \label{eq:support-scale}
\end{align}
The prior interval mixing the same contacts has width
\begin{equation}
y_{2m,m}-x_{2m,m}=\frac3{8m^2}+\frac{15t}{64m^4}+O(m^{-6}), \label{eq:prior-width-scale}
\end{equation}
while the two admissible endpoint windows have widths
\begin{align}
x_{2m,m}-\tau_{2m+1}&=\frac1{16m^3}-\frac{3t+2}{64m^4}+O(m^{-5}), \notag\\
\tau_m-y_{2m,m}&=\frac1{2m^3}+\frac{8-3t}{16m^4}+O(m^{-5}). \label{eq:endpoint-window-scale}
\end{align}
At the midpoint prior, the gain over full revelation is $17/(1536m^4)+O(m^{-5})$, and the gain over no information is $O(m^{-4})$.
\end{proposition}

\begin{proof}
Taylor expansion of \eqref{eq:xnm}--\eqref{eq:ynm} and $\tau_{2m+1},\tau_m$ yields the first four displays. Let $p_m=(\mu_{0m}-L_m)/(H_m-L_m)=1/2+O(m^{-1})$. The tangent's square gaps at the endpoints give
\[
G_m^F=(1-p_m)\frac{4m^2-t}{3}(x_{2m,m}-L_m)^2
+p_m\frac{m^2-t}{3}(H_m-y_{2m,m})^2.
\]
Here $x_{2m,m}-L_m=1/(32m^3)+O(m^{-4})$ and $H_m-y_{2m,m}=1/(4m^3)+O(m^{-4})$, giving the stated coefficient. The uniform $O(m^{-4})$ comparison with no information follows from the branch identity and tangent expansion in Appendix~\ref{app:scale-proof}.
\end{proof}

The leading absolute normalized loss at this prior is $5/(48m^2)$. Thus the gain over full revelation is only $17/(160m^2)+o(m^{-2})$ as a fraction of that leading loss. Arbitrarily many skipped regimes is a geometric possibility, not evidence of a large policy gain in the limiting construction.

\subsection{Scale calculations for the extensive-skipping construction}\label{app:scale-proof}

For fixed $t$ and $n=2m$, the contacts can be written
\[
x_{2m,m}=\frac1{8m^2}
\sqrt{\frac{1-t/m^2}{1-t/(4m^2)}},
\qquad
y_{2m,m}=\frac1{2m^2}
\sqrt{\frac{1-t/(4m^2)}{1-t/m^2}}.
\]
Expanding these expressions gives \eqref{eq:contact-scale}. The thresholds satisfy
\[
\tau_{2m+1}=\frac1{8m^2}-\frac1{16m^3}+\frac1{32m^4}+O(m^{-5}),
\qquad
\tau_m=\frac1{2m^2}+\frac1{2m^3}+\frac1{2m^4}+O(m^{-5}),
\]
which gives \eqref{eq:support-scale}--\eqref{eq:endpoint-window-scale} by substitution and averaging. The prior mean is
\[
\mu_{0m}=\frac{x_{2m,m}+y_{2m,m}}2
=\frac5{16m^2}+\frac{9t}{128m^4}+O(m^{-6}).
\]
The square gaps at $L_m$ and $H_m$ imply the full-revelation comparison in Proposition~\ref{prop:skipping-scale}. To bound the gain over no information, note the exact branch identity
\begin{equation}
W_j(\mu)=-\frac\mu3+\frac{t\mu^2}{3}
-\frac{(2j^2\mu-1)^2}{12j^2}. \label{eq:affine-limit-identity}
\end{equation}
On a selected regime with $\mu=O(m^{-2})$ in the support construction, $j=\Theta(m)$, and the last two terms are $O(m^{-4})$. The tangent has the uniform expansion
\[
T_{2m,m}(\mu)
=-\frac\mu3+\frac{5t}{24m^2}\mu
-\frac{t}{48m^4}+O(m^{-6})
\quad\text{for }\mu=O(m^{-2}),
\]
so $T_{2m,m}(\mu_{0m})-W(\mu_{0m})=O(m^{-4})$. Also $-\mu_{0m}/3=-5/(48m^2)+O(m^{-4})$, as used for the relative-gain calculation.

\subsection{Exact values in the main example}\label{app:example-exact}

The prior high-state probability is $p=551/2351$. Bayes plausibility yields
\begin{equation}
\Prob(\mu_z=x)=\frac{553}{1000},
\qquad
\Prob(\mu_z=y)=\frac{447}{1000}. \label{eq:extended-example-weights}
\end{equation}
The posterior high-state probabilities are $q_x=132/16457$ and $q_y=3628/7053$. An exact implementing experiment is
\begin{equation}
\begin{array}{c|cc}
&x&y\\ \hline
B=L &\dfrac{51587}{72000}&\dfrac{20413}{72000}\\[0.8em]
B=H &\dfrac{2607}{137750}&\dfrac{135143}{137750}
\end{array}. \label{eq:extended-example-signal}
\end{equation}
These probabilities follow from \eqref{eq:binary-implementation} and sum to one in each state.

For the normalized common-study objective, the three benchmark values are
\begin{align*}
\text{no information:}\quad&W_3(\mu_0)=-\frac{8669}{337500}\approx-0.02568593,\\
\text{optimal study:}\quad&T(\mu_0)=-\frac{4357}{192000}\approx-0.02269271,\\
\text{full revelation:}\quad&(1-p)W_5(L)+pW_1(H)=-\frac{855126733}{35265000000}\approx-0.02424860.
\end{align*}
The gain over full revelation is $146315663/94040000000\approx0.00155589$, and the gain over no information is $129307/43200000\approx0.00299322$. Relative to full revelation, the communication effect is $433943471/197484000000\approx0.00219736$ and the forecast effect is $-538837/840000000\approx-0.00064147$.

To verify support beyond the theorem's window: on $[L,\tau_5]$ the selected branch is $W_5$, with $T(L)-W_5(L)=1227/500000000>0$ and $(T-W_5)'(L)=3273/25000>0$; the gap is strictly convex, so it stays positive up to $\tau_5$. On $[\tau_2,H]$ the branch is $W_1$, with $T(H)-W_1(H)=10609/1600000>0$ and $(T-W_1)'(\mu)=-\frac{21}{80}-\frac\mu5<0$, so the gap is positive throughout. Lemma~\ref{lem:common-tangent} covers $W_4,W_3,W_2$.

The example's low-support margin is narrow. For the fixed $c$ and contacts in \eqref{eq:example-tangent}, the exact connected range of endpoints containing both contacts for which this tangent has only those two equality points is $L_*<L<x<y<H<H_*$, where
\[
L_* = \frac{105+\sqrt{2730}}{6320}\approx0.02488123,
\qquad
H_* = \frac{5\sqrt{26}-21}{16}\approx0.28094360.
\]
The displayed low state $L=0.0249$ clears $L_*$ by about $1.88\times10^{-5}$; lowering $L$ by one percent makes this same tangent fail to support the value. Strict support, and hence the local density-perturbation result in Proposition~\ref{prop:density-robustness}, holds on an open set, but these numbers do not imply a large quantitative robustness radius.

The narrower support $[13/500,3/20]$ with the same $c$ and $\mu_0$ also satisfies Theorem~\ref{thm:global-skipping}, with the same contacts. Full revelation there induces four and two messages, and the normalized gain over it is only $2111/248000000$. Figure~\ref{fig:overview} and Section~\ref{sec:retained-information} use this support.

\section{Details for Section~\ref{sec:retained-information}}\label{app:retained-details}

\subsection{Which public reports can be produced}

Write $\nu_r=\E[B\mid r]$ and $\mu_z=\E[B\mid z]$.
The raw experiment also restricts which public posterior distributions can be generated. Since
\begin{equation}
\mu_z=\E[\nu_r\mid z], \label{eq:posterior-martingale}
\end{equation}
the law of $\mu_z$ must be a mean-preserving contraction of the law of $\nu_r$:
\begin{equation}
\mathcal L(\mu_z)\preceq_{\mathrm{cx}}\mathcal L(\nu_r). \label{eq:convex-order-constraint}
\end{equation}
Equivalently, the two variables have the same mean and
\[
\E[\phi(\mu_z)]\leq\E[\phi(\nu_r)]
\]
for every convex function $\phi$ for which the expectations exist. Conversely, any distribution satisfying \eqref{eq:convex-order-constraint} admits a martingale coupling with $\nu_r$ and can be implemented by randomizing the public report as a function of the raw posterior. If $r$ fully reveals binary $B$, then $\nu_r=B\in\{L,H\}$; this endpoint distribution dominates every distribution on $[L,H]$ with mean $\mu_0$ in convex order. Thus every Bayes-plausible distribution of public posterior means is implementable.

\subsection{The exact retained-information example}

Let $T^R_{n,m}$ denote the line \eqref{eq:Tnm} evaluated at $t=1$,
\begin{equation}
T^R_{n,m}(\mu)\equiv-\frac1{12m^2n^2}-\frac{\sqrt{(n^2-1)(m^2-1)}}{3mn}\mu. \label{eq:retained-tangent}
\end{equation}

Return to
\[
L=\frac{13}{500},
\qquad
H=\frac3{20},
\qquad
\mu_0=\frac2{25},
\qquad
n=4,
\qquad
m=2.
\]
Equations \eqref{eq:retained-x}--\eqref{eq:retained-y} give
\[
x^R=\frac1{16\sqrt5}=\frac{\sqrt5}{80},
\qquad
y^R=\frac{\sqrt5}{16}.
\]
They satisfy
\[
\tau_5=\frac1{40}
<L
<x^R
<\tau_4=\frac1{24}
\]
and
\[
\tau_3=\frac1{12}
<y^R
<H
<\tau_2=\frac14.
\]
The public posterior weights are
\begin{equation}
\lambda_x
=\frac{125-32\sqrt5}{100},
\qquad
\lambda_y
=\frac{32\sqrt5-25}{100}. \label{eq:retained-example-weights}
\end{equation}
Here $p=27/62$, and the posterior high-state probabilities are
\[
q_x=\frac{25\sqrt5-52}{248},
\qquad
q_y=\frac{125\sqrt5-52}{248}.
\]
An exact committed disclosure rule is
\begin{equation}
\begin{array}{c|cc}
&x^R&y^R\\ \hline
B=L
&\dfrac{1660-509\sqrt5}{560}
&\dfrac{509\sqrt5-1100}{560}\\[0.9em]
B=H
&\dfrac{4789\sqrt5-10500}{10800}
&\dfrac{21300-4789\sqrt5}{10800}
\end{array}. \label{eq:retained-example-signal}
\end{equation}

Holding full private information fixed, no public disclosure gives
\[
V_3\left(\frac2{25}\right)
=-\frac{1777}{67500}
\approx-0.02632593.
\]
The two-contact disclosure attains
\[
T^R_{4,2}\left(\frac2{25}\right)
=-\frac1{768}-\frac{\sqrt5}{100}
\approx-0.02366276,
\]
whereas full public disclosure gives
\[
\frac{35}{62}V_4\left(\frac{13}{500}\right)
+\frac{27}{62}V_2\left(\frac3{20}\right)
=-\frac{176471}{7440000}
\approx-0.02371922.
\]
Thus
\[
\text{no public disclosure}
<\text{full public disclosure}
<\text{two-contact disclosure}.
\]
The gains of the two-contact disclosure over no disclosure and full disclosure are, respectively,
\[
\frac{108103}{4320000}-\frac{\sqrt5}{100}
\approx0.00266316
\]
and
\[
\frac{111189}{4960000}-\frac{\sqrt5}{100}
\approx0.00005646.
\]

Here public garbling improves communication without reducing the principal's information at the action stage: the raw signal fully reveals $B$ under every disclosure rule.

\subsection{The full-revelation threshold}\label{app:threshold}

Let $\ell_V$ denote the affine chord through $(L,V(L))$ and $(H,V(H))$. The endpoint chord of $W_c=V+c\mu^2$ is
\begin{equation}
\Lambda_c(\mu)=\ell_V(\mu)+c[(L+H)\mu-LH], \label{eq:common-endpoint-chord}
\end{equation}
and its gap from $W_c$ is
\begin{equation}
\Lambda_c(\mu)-W_c(\mu)
=\ell_V(\mu)-V(\mu)+c(\mu-L)(H-\mu). \label{eq:common-chord-gap}
\end{equation}

\begin{proposition}[A full-revelation threshold under a common study]\label{prop:full-threshold}
For fixed $0<L<H$, define
\begin{equation}
c_F(L,H)=\max\left\{0,\sup_{\mu\in(L,H)}
\frac{V(\mu)-\ell_V(\mu)}{(\mu-L)(H-\mu)}\right\}. \label{eq:full-revelation-threshold}
\end{equation}
This threshold is finite. For every $c>c_F(L,H)$ and every interior prior, full revelation is the unique optimal common-study posterior-mean law. At $c=c_F(L,H)$, full revelation is optimal, though it need not be unique. With costless full private acquisition, the disclosure objective is independent of $c$; hence an interior retained-information optimum can coexist with unique common-study full revelation at the same primitives.
\end{proposition}

\begin{proof}
On a fixed compact support with $L>0$, only finitely many selected branches occur. Their values are continuous and piecewise smooth with bounded one-sided slopes, so the ratio in \eqref{eq:full-revelation-threshold} is bounded above, including near both endpoints. Equation~\eqref{eq:common-chord-gap} shows strict interior domination by $\Lambda_c$ for $c>c_F$. Every feasible posterior law has the same mean and yields expected value weakly below the chord. Equality is possible only at $L$ and $H$; their weights are uniquely fixed by Bayes plausibility. At equality $c=c_F$ the chord still dominates but can have additional contacts. The retained-information policy comparison follows from Proposition~\ref{prop:separation}.
\end{proof}

For the support and prior in the preceding retained-information example,
\[
L=\frac{13}{500},\qquad H=\frac3{20},\qquad\mu_0=\frac2{25},
\]
the exact threshold is
\begin{equation}
c_F=\frac{2505}{15376}\approx0.162916. \label{eq:exact-full-threshold}
\end{equation}
Appendix~\ref{app:architecture-threshold} verifies the chord gaps branch by branch. Take the same actual preference parameter $c=1/5>c_F$ in both architectures. The common-study optimum is uniquely full revelation, with endpoint weights $35/62,27/62$ and normalized value
\begin{equation}
\frac{35}{62}W_4\!\left(\frac{13}{500}\right)
+\frac{27}{62}W_2\!\left(\frac3{20}\right)
=-\frac{4033079}{186000000}\approx-0.02168322. \label{eq:policy-difference-common}
\end{equation}
Under full private retention, the unique optimal public disclosure remains the two-contact law in \eqref{eq:retained-example-weights}, with posteriors $x^R=\sqrt5/80$ and $y^R=\sqrt5/16$. Its disclosure objective is
\[
\E[V(\mu_z^R)]=-\frac1{768}-\frac{\sqrt5}{100}\approx-0.02366276,
\]
whereas full public disclosure, holding full private acquisition fixed, gives $-176471/7440000\approx-0.02371922$. The two objective normalizations contain different residual-risk constants and should be compared \emph{within} an architecture. Indeed, if the retained-information two-contact posterior law were evaluated under the common-study objective at $c=1/5$, it would yield $-1/480-11\sqrt5/1250\approx-0.02176073$, below the common-study value under full revelation. Even at $c=1/10$, both architectures choose interior designs on this support, but their optimal contact posteriors differ.

\subsection{Exact full-revelation threshold in the architecture example}\label{app:architecture-threshold}

On $[13/500,3/20]$, the communication-value chord is
\[
\ell_V(\mu)=-\frac{1939}{1488000}-\frac{6949}{24800}\mu.
\]
Set $c_*=2505/15376$. Direct substitution into \eqref{eq:common-endpoint-chord} gives the chord gaps on the three selected regimes:
\begin{align*}
\Lambda_{c_*}(\mu)-W_{4,c_*}(\mu)
&=\frac{74375}{15376}\left(\mu-\frac{13}{500}\right)^2,\\
\Lambda_{c_*}(\mu)-W_{3,c_*}(\mu)
&=\frac{115493}{46128}\mu^2
-\frac{7735}{30752}\mu
+\frac{1215697}{166060800},\\
\Lambda_{c_*}(\mu)-W_{2,c_*}(\mu)
&=\frac{12871}{15376}\left(\mu-\frac3{20}\right)
\left(\mu-\frac{38737}{257420}\right).
\end{align*}
The middle quadratic has positive leading coefficient and negative discriminant. In the last expression the second root exceeds $H=3/20$, so the product is positive for $\mu<H$ in its regime. Equality occurs only at the endpoints. Thus $c_F\leq c_*$. If $c<c_*$, then
\[
\Lambda_c(\mu)-W_c(\mu)
=\frac{74375}{15376}(\mu-L)^2
-(c_*-c)(\mu-L)(H-\mu)
\]
is negative for $\mu>L$ sufficiently close to $L$ in the four-message regime. Hence the endpoint chord fails to support $W_c$ and $c_F\geq c_*$. This proves \eqref{eq:exact-full-threshold}, including the threshold equality case: the endpoint chord touches only $L,H$ in this particular example.

\section{The Zero-Conflict Benchmark in Full}\label{app:zero-details}

This appendix gives the full characterization behind Proposition~\ref{prop:zero-summary}. Maintain the common study and specialize to
\[
B\in\{0,H\},
\qquad
\Prob(B=H)=p\in(0,1),
\qquad
\mu_0=pH,
\qquad
0\leq c<1.
\]
Every distribution of posterior means on $[0,H]$ with mean $\mu_0$ is implementable. The normalized problem is therefore
\begin{equation}
\sup_{\substack{\mu_z\in[0,H]\\ \E[\mu_z]=\mu_0}}
\E[W(\mu_z)]
=\operatorname{cav}_{[0,H]}W(\mu_0),
\qquad
W(0)=0. \label{eq:zero-baseline-problem}
\end{equation}
The principal's original payoff differs by the prior-fixed term $-cH\mu_0$.

The origin changes the global geometry. The internal common tangent $T_{n,m}$ from Section~\ref{sec:positive-L} has
\[
T_{n,m}(0)=-\frac{t}{12m^2n^2}<0=W(0).
\]
It therefore cannot support the selected value on a domain containing zero. The nonadjacent tangencies used in Theorem~\ref{thm:global-skipping} remain valid branch comparisons, but they cannot appear in the global envelope of the present benchmark.

\subsection{Origin contacts and transition points}

Retain $t=1+3c$ and, for every $k\geq2$, write
\begin{equation}
A_k\equiv\frac1{12k^2},
\qquad
d_k\equiv\frac{k^2-t}{3},
\qquad
W_k(\mu)=-A_k-d_k\mu^2. \label{eq:zero-branch-notation}
\end{equation}
Because $c<1$, every informative branch is strictly concave. Define
\begin{equation}
r_k\equiv\frac1{2k\sqrt{k^2-t}},
\qquad
\beta_k\equiv\frac{\sqrt{k^2-t}}{3k},
\qquad
\ell_k(\mu)\equiv-\beta_k\mu. \label{eq:origin-contact-definitions}
\end{equation}
Indeed,
\[
\frac{W_k(u)}u=-\frac{A_k}{u}-d_ku
\]
is uniquely maximized at $u=r_k$, where it equals $-\beta_k$. Thus $\ell_k$ passes through the origin and is tangent to $W_k$ at $r_k$.

The contact always satisfies $r_k>\tau_{k+1}$, while
\begin{equation}
r_k<\tau_k
\quad\Longleftrightarrow\quad
t<2k-1. \label{eq:origin-contact-feasibility}
\end{equation}
For $0\leq c<1$, the latter condition is automatic for $k\geq3$. It can fail only for $k=2$, in which case $r_2\geq\tau_2=1/4$ iff $c\geq2/3$.

For $0<H<1/4$, choose the unique $n\geq2$ with
\begin{equation}
H\in[\tau_{n+1},\tau_n). \label{eq:H-in-n-regime}
\end{equation}
At $H=1/4$, set $n=2$. Thus a lower communication threshold is assigned to the coarser adjacent regime, consistently with the selected intervals in \eqref{eq:regime-interval}; continuity gives the same envelope as the finer-side limit. Let
\begin{equation}
\bar r_n\equiv\min\{r_n,\tau_n\} \label{eq:rbar-definition}
\end{equation}
and define
\begin{equation}
h_n
\equiv
\frac{
\beta_{n+1}-
\sqrt{\beta_{n+1}^2-4d_nA_n}
}{2d_n}. \label{eq:hn-definition}
\end{equation}
The point $h_n$ is the smaller solution of
\begin{equation}
\frac{W_n(h_n)}{h_n}=-\beta_{n+1}. \label{eq:hn-ratio-equality}
\end{equation}
Since $4d_nA_n=\beta_n^2$ and $\beta_{n+1}>\beta_n$, the square root in \eqref{eq:hn-definition} is real. Direct comparison gives
\begin{equation}
\tau_{n+1}<h_n<\bar r_n\leq\tau_n,
\qquad
\tau_{n+2}<r_{n+1}<\tau_{n+1}. \label{eq:low-ordering}
\end{equation}

When $H\geq\tau_{n+1}$, define
\begin{equation}
\xi_n(H)
\equiv
H-
\sqrt{
\frac{2n+1}{(n+1)^2-t}
\left(H^2-\tau_{n+1}^2\right)
}. \label{eq:xi-c-definition}
\end{equation}
The line
\begin{equation}
T_{n,H}(\mu)
\equiv
W_{n+1}\bigl(\xi_n(H)\bigr)
+W_{n+1}'\bigl(\xi_n(H)\bigr)
\bigl(\mu-\xi_n(H)\bigr) \label{eq:endpoint-tangent-low}
\end{equation}
is tangent to $W_{n+1}$ at $\xi_n(H)$ and, by construction, satisfies
\begin{equation}
T_{n,H}(H)=W_n(H). \label{eq:endpoint-tangent-low-contact}
\end{equation}

For the babbling range, suppose first that $0\leq c<2/3$ and define
\begin{equation}
r\equiv\frac1{4\sqrt{3(1-c)}},
\qquad
\beta\equiv\sqrt{\frac{1-c}{12}}, \label{eq:high-origin-contact}
\end{equation}
and
\begin{equation}
H_*
\equiv
\frac1{
\sqrt3\bigl(\sqrt{1+3c}+\sqrt{1-c}\bigr)
}. \label{eq:Hstar-definition}
\end{equation}
Then $r<1/4<H_*$. For $1/4<H<H_*$, let
\begin{equation}
x(H)
\equiv
H-
\sqrt{\frac{H^2-1/16}{1-c}} \label{eq:high-contact}
\end{equation}
and define
\begin{equation}
S_H(\mu)
\equiv
W_2\bigl(x(H)\bigr)
+W_2'\bigl(x(H)\bigr)\bigl(\mu-x(H)\bigr). \label{eq:high-endpoint-tangent}
\end{equation}
This line is tangent to $W_2$ at $x(H)$ and passes through $(H,W_1(H))$.

\begin{theorem}[Closed-form zero-baseline benchmark]\label{thm:zero-baseline}
Suppose $B\in\{0,H\}$ and $0\leq c<1$.

\smallskip
\noindent\emph{A. No babbling on the posterior domain: $0<H\leq1/4$.}
Choose $n$ as in \eqref{eq:H-in-n-regime}.

\begin{enumerate}[label=\textup{(A\arabic*)},leftmargin=2.4em]
\item If $\tau_{n+1}\leq H<h_n$, then
\begin{equation}
\operatorname{cav}W(\mu)=
\begin{cases}
-\beta_{n+1}\mu,
&0\leq\mu\leq r_{n+1},\\[2pt]
W_{n+1}(\mu),
&r_{n+1}\leq\mu\leq\xi_n(H),\\[2pt]
T_{n,H}(\mu),
&\xi_n(H)\leq\mu\leq H.
\end{cases} \label{eq:low-envelope-bridge}
\end{equation}

\item If $h_n<H\leq\bar r_n$, then full revelation is supported by the origin-to-endpoint chord:
\begin{equation}
\operatorname{cav}W(\mu)
=\frac{\mu}{H}W_n(H). \label{eq:low-envelope-full}
\end{equation}

\item If $\bar r_n<H\leq\tau_n$, a case that is nonempty only when $r_n<\tau_n$, then
\begin{equation}
\operatorname{cav}W(\mu)=
\begin{cases}
-\beta_n\mu,
&0\leq\mu\leq r_n,\\[2pt]
W_n(\mu),
&r_n\leq\mu\leq H.
\end{cases} \label{eq:low-envelope-origin-n}
\end{equation}
\end{enumerate}

At $H=h_n$, the envelope is the single chord $-\beta_{n+1}\mu$ and its equality set with $W$ is $\{0,r_{n+1},H\}$. If $r_n<\tau_n$, the formulas in \textup{(A2)} and \textup{(A3)} agree at their common boundary $H=\bar r_n=r_n$. If $r_n\geq\tau_n$, case \textup{(A3)} is empty and \textup{(A2)} governs through $H=\bar r_n=\tau_n$. Communication-threshold endpoints are interpreted using continuity of the selected value.

\smallskip
\noindent\emph{B. The posterior domain contains babbling: $H>1/4$.}

\begin{enumerate}[label=\textup{(B\arabic*)},leftmargin=2.4em]
\item If $0\leq c<2/3$ and $1/4<H<H_*$, then
\begin{equation}
\operatorname{cav}W(\mu)=
\begin{cases}
-\beta\mu,
&0\leq\mu\leq r,\\[2pt]
W_2(\mu),
&r\leq\mu\leq x(H),\\[2pt]
S_H(\mu),
&x(H)\leq\mu\leq H.
\end{cases} \label{eq:high-envelope-bridge}
\end{equation}

\item If $0\leq c<2/3$ and $H>H_*$, or if $2/3\leq c<1$ and $H>1/4$, then
\begin{equation}
\operatorname{cav}W(\mu)
=\frac{\mu}{H}W_1(H). \label{eq:high-envelope-full}
\end{equation}
\end{enumerate}

At $H=H_*$, the envelope is the chord $-\beta\mu$ and its equality set with $W$ is $\{0,r,H\}$.
\end{theorem}

\begin{proof}
The proof has two steps. First, the best line leaving the origin is found by maximizing $W(u)/u$. When $H$ lies in the $n$-message regime, every branch finer than $n+1$ has a strictly worse maximal ratio. The comparison therefore reduces to the stationary contact $r_{n+1}$ and the best feasible point on branch $n$. Their ranking changes at $h_n$ and $\bar r_n$, yielding cases \textup{(A1)}--\textup{(A3)}. Second, each remaining exposed branch is joined to the endpoint by the unique tangent in \eqref{eq:endpoint-tangent-low}; concavity and a slope comparison establish global domination.

When $H>1/4$, the informative candidate reduces to the two-message branch, to be compared with the babbling endpoint. For $c<2/3$, the two-message stationary contact lies inside its regime and its ranking against the endpoint changes at $H_*$. Below that value, the envelope follows the two-message branch and then its tangent to the endpoint. Above it, the origin-to-endpoint chord dominates globally. If $c\geq2/3$, the stationary two-message contact lies beyond its regime and the endpoint chord dominates for every $H>1/4$. Appendix~\ref{app:zero-proof} supplies the finite-ratio reduction, contact ordering, and global-support calculations.
\end{proof}

The theorem shows that, when $H\leq1/4$, only the regime containing $H$ and its immediate finer neighbor can be exposed. All still-finer regimes are eliminated by the origin tangent in \eqref{eq:finite-origin-comparison}; no coarser selected regime occurs on the domain. Global nonadjacent internal tangencies are therefore absent even though the corresponding extrapolated branch tangencies exist.

\subsection{Optimal experiments}

The envelope immediately gives the optimal posterior distributions. The descriptions below are for generic interior priors; at a contact, the corresponding probability becomes zero and the experiment degenerates continuously.

In case \textup{(A1)}, the optimal support is
\begin{equation}
\begin{cases}
\{0,r_{n+1}\},
&0<\mu_0<r_{n+1},\\
\{\mu_0\},
&r_{n+1}<\mu_0<\xi_n(H),\\
\{\xi_n(H),H\},
&\xi_n(H)<\mu_0<H.
\end{cases} \label{eq:A1-supports}
\end{equation}
The nondegenerate weights are
\begin{align}
\Prob(\mu_z=r_{n+1})
&=\frac{\mu_0}{r_{n+1}},
&
\Prob(\mu_z=0)
&=1-\frac{\mu_0}{r_{n+1}}, \label{eq:origin-weights-low}\\
\Prob(\mu_z=H)
&=\frac{\mu_0-\xi_n(H)}{H-\xi_n(H)},
&
\Prob\bigl(\mu_z=\xi_n(H)\bigr)
&=\frac{H-\mu_0}{H-\xi_n(H)}. \label{eq:endpoint-weights-low}
\end{align}
Thus the middle region uses no information, while the two affine regions use partial information.

In case \textup{(A2)}, full revelation is optimal:
\begin{equation}
\Prob(\mu_z=H)=\frac{\mu_0}{H}=p,
\qquad
\Prob(\mu_z=0)=1-p. \label{eq:zero-full-revelation-weights}
\end{equation}
In case \textup{(A3)}, priors below $r_n$ use $\{0,r_n\}$ with the weights in \eqref{eq:origin-weights-low}, replacing $r_{n+1}$ by $r_n$; priors above $r_n$ use no information.

In the three-piece babbling case \textup{(B1)}, the optimal support is
\begin{equation}
\begin{cases}
\{0,r\},
&0<\mu_0<r,\\
\{\mu_0\},
&r<\mu_0<x(H),\\
\{x(H),H\},
&x(H)<\mu_0<H.
\end{cases} \label{eq:B1-supports}
\end{equation}
The weights are respectively
\begin{equation}
\Prob(\mu_z=r)=\frac{\mu_0}{r},
\qquad
\Prob(\mu_z=0)=1-\frac{\mu_0}{r}, \label{eq:high-origin-weights}
\end{equation}
and
\begin{equation}
\Prob(\mu_z=H)=\frac{\mu_0-x(H)}{H-x(H)},
\qquad
\Prob\bigl(\mu_z=x(H)\bigr)=\frac{H-\mu_0}{H-x(H)}. \label{eq:high-endpoint-weights}
\end{equation}
Case \textup{(B2)} uses full revelation with \eqref{eq:zero-full-revelation-weights}.

Away from transition values, these posterior distributions are unique up to splitting or relabelling signals. Strict concavity makes no information unique on every exposed branch, while an affine segment has only its two stated contact posteriors in its equality set. At $H=h_n$, any Bayes-plausible distribution supported on $\{0,r_{n+1},H\}$ is optimal. At $H=H_*$, the corresponding equality set is $\{0,r,H\}$. If $r_n<\tau_n$, the \textup{(A2)} and \textup{(A3)} formulas coincide at $H=\bar r_n=r_n$, and the equality set of the chord is $\{0,H\}$. If $r_n\geq\tau_n$, case \textup{(A3)} is empty, so $H=\bar r_n=\tau_n$ creates no separate case transition. At communication thresholds, the posterior law is unaffected, although the adjacent equilibrium labels coincide because one cell has zero length.

No-information disclosure is never optimal for a prior strictly inside the babbling interval $(1/4,H)$. In case \textup{(B1)}, every such prior lies on the strict endpoint-tangent segment. In case \textup{(B2)}, it lies on the strict full-revelation chord.

\subsection{Boundary consistency and scope}

The two parts of Theorem~\ref{thm:zero-baseline} agree at $H=1/4$. If $c<2/3$, then $r_2<1/4$; as $H\downarrow1/4$ in case \textup{(B1)}, $x(H)\uparrow1/4$ and the endpoint-tangent segment disappears. The remaining envelope is $-\beta_2\mu$ up to $r_2$ and $W_2$ thereafter, exactly case \textup{(A3)}. If $c\geq2/3$, then $r_2\geq1/4$ and the origin-to-$H$ chord is already optimal at $H=1/4$, matching cases \textup{(A2)} and \textup{(B2)}.

As $c\downarrow0$, $t\downarrow1$, the contacts and transition values converge to the original communication-only benchmark. In particular,
\[
r\longrightarrow\frac1{4\sqrt3},
\qquad
H_*\longrightarrow\frac1{2\sqrt3}.
\]
For $c>0$, however, the term $c\Var(\mu_z)$ is the study's direct decision value. Full revelation maximizes that component, so whenever it is optimal its value cannot be described as purely strategic.

\subsection{Proof details for Theorem~\ref{thm:zero-baseline}}\label{app:zero-proof}

Suppose first that $H$ satisfies \eqref{eq:H-in-n-regime}. Every selected branch finer than $n+1$ obeys
\[
\frac{W_k(u)}u\leq-\beta_k<-\beta_{n+1}.
\]
The $(n+1)$-message branch reaches $-\beta_{n+1}$ at $r_{n+1}$. On the $n$-message branch, the ratio increases up to $r_n$ and decreases afterward. Hence
\begin{equation}
\sup_{0<u\leq H}\frac{W(u)}u
=\max\left\{
-\beta_{n+1},
\frac{W_n(u_n)}{u_n}
\right\},
\qquad
u_n\equiv\min\{H,r_n\}. \label{eq:finite-origin-comparison}
\end{equation}
This reduces the infinite collection of informative regimes to two candidates and rules out exposure of every regime finer than $n+1$.

On $[\tau_{n+1},\bar r_n]$, the ratio $W_n(H)/H$ is strictly increasing and crosses $-\beta_{n+1}$ at $h_n$. The origin support is therefore $r_{n+1}$ when $H<h_n$, the endpoint $H$ when $h_n<H\leq\bar r_n$, and $r_n$ when $H>\bar r_n$.

In the first case, solving for a tangent to $W_{n+1}$ that passes through $(H,W_n(H))$ gives \eqref{eq:xi-c-definition}. As $H$ rises from $\tau_{n+1}$ to $h_n$, the contact $\xi_n(H)$ falls from $\tau_{n+1}$ to $r_{n+1}$. Strict concavity makes the tangent dominate $W_{n+1}$. On the $n$-message interval, direct substitution gives
\[
d_nH<d_{n+1}\xi_n(H).
\]
Thus $T_{n,H}-W_n$ is decreasing on $[\tau_{n+1},H]$ and reaches zero only at $H$, proving \eqref{eq:low-envelope-bridge}. In the second case, the endpoint ratio dominates every finer-branch ratio and every earlier point of $W_n$, which proves \eqref{eq:low-envelope-full}. In the third, $\ell_n$ dominates $W_n$ by concavity and every finer selected branch because $\beta_n<\beta_k$ for $k>n$, proving \eqref{eq:low-envelope-origin-n}. The defining equation for $h_n$ gives the three-point equality set at the transition.

Now let $H>1/4$. If $c<2/3$, the two-message origin contact $r$ lies strictly inside its selected regime. The babbling endpoint ratio
\begin{equation}
\frac{W_1(H)}H=-\frac1{12H}+cH \label{eq:babbling-endpoint-ratio}
\end{equation}
is strictly increasing and equals $-\beta$ exactly at $H_*$. For $H<H_*$, the endpoint-tangency equation gives \eqref{eq:high-contact}, with $r<x(H)<1/4$. The resulting line dominates $W_2$ by concavity. It also dominates the entire babbling branch because $S_H-W_1$ is concave on $[1/4,H]$, is nonnegative at $1/4$, and vanishes at $H$.

For $H>H_*$, the endpoint ratio exceeds the ratio at every informative posterior and every earlier babbling posterior, so the origin-to-endpoint chord dominates globally. At equality it also touches $W_2$ at $r$, producing equality set $\{0,r,H\}$. Finally, if $c\geq2/3$, then $r_2\geq1/4$. The informative ratio is maximized at $1/4$, while the babbling ratio is increasing; the endpoint chord therefore dominates for every $H>1/4$.

\section{Local robustness of the supporting geometry}\label{app:geometric-robustness}

\begin{proposition}[Regular perturbations of a strict two-contact support]\label{prop:geometric-robustness}
Let a continuous selected value $W^0$ have finitely many $C^2$ branches on a compact interval $[L,H]$. Suppose an affine $T^0$ is tangent to its $n$- and $m$-message branches at $L<x^0<y^0<H$, the two branches have strictly negative second derivatives near their contacts, and $T^0>W^0$ everywhere else on $[L,H]$. Suppose contacts and support endpoints lie strictly away from the relevant thresholds. Under sufficiently small $C^2$ perturbations of the contact branches near $x^0,y^0$, together with uniformly small perturbations of the selected value on the rest of $[L,H]$ and of threshold locations, preserving continuity and the local regime order, there is a nearby common tangent $T^\varepsilon$ with exactly two equality points $x^\varepsilon,y^\varepsilon$ on $[L,H]$. For a binary endpoint prior whose mean lies between them, the unique optimal posterior law remains the corresponding two-contact law.
\end{proposition}

\begin{proof}
Writing the common slope as $s$, the contact equations are
\[
(W_n^\varepsilon)'(x)-s=0,\qquad
(W_m^\varepsilon)'(y)-s=0,\qquad
W_n^\varepsilon(x)-sx-W_m^\varepsilon(y)+sy=0.
\]
At the original solution, their Jacobian with respect to $(x,y,s)$ is
\[
\begin{pmatrix}
(W_n^0)''(x^0)&0&-1\\
0&(W_m^0)''(y^0)&-1\\
0&0&y^0-x^0
\end{pmatrix},
\]
whose determinant is nonzero. The implicit-function theorem gives unique nearby contacts and a common tangent. Strict concavity preserves the positive gap in punctured neighborhoods of those contacts. Outside the neighborhoods, the original gap has a positive minimum on a compact set, which uniform closeness preserves, including at shifted thresholds. Thus the perturbed line is a global majorant with exactly two contacts. Binary Bayes plausibility and the equality set establish uniqueness.
\end{proof}

The preceding proposition is about abstract continuation values. For the maintained quadratic game, the next result verifies its relevant local conclusions directly from a perturbation of the manager's type density. In particular, it does not assume that off-path feasibility thresholds have an unchanged order.

\begin{lemma}[Uniform upper control of nearby partition payoffs]\label{lem:partition-upper-control}
Fix $0<L<H$ and a finite $c\geq0$. Let $f$ be a strictly positive, normalized density on $[0,1]$ with $\|f-1\|_{C^2}$ small, independent of the residual state. For $\mu\in[L,H]$, let $M_f(\mu)$ be the supremum of normalized principal payoffs $V_j^f(\mu)+c\mu^2$ over all feasible monotone pure-strategy partition equilibria, including babbling. Let $W^0$ be the selected finest-partition value for the uniform density. Then the number of cells in any feasible equilibrium is bounded uniformly over $\mu$ and nearby $f$, and
\begin{equation}
\sup_{\mu\in[L,H]}\bigl(M_f(\mu)-W^0(\mu)\bigr)_+
\longrightarrow0\quad\text{as }\|f-1\|_{C^2}\longrightarrow0. \label{eq:upper-partition-control}
\end{equation}
This conclusion includes posteriors at which a cell disappears in the uniform-type limit.
\end{lemma}

\begin{proof}
Consider a feasible $j$-cell partition $0=b_0<b_1<\cdots<b_j=1$, with lengths $\ell_i=b_i-b_{i-1}$ and conditional means
\[
q_i(f)=\frac{\int_{b_{i-1}}^{b_i}\theta f(\theta)\,d\theta}
{\int_{b_{i-1}}^{b_i}f(\theta)\,d\theta}.
\]
Here $V_j^f(\mu)=-\sum_i\int_{b_{i-1}}^{b_i}(\theta-q_i(f))^2f(\theta)\,d\theta$, the within-cell communication payoff. 
Put $\eta=\|f-1\|_\infty<1$ and $e_i=q_i(f)-(b_{i-1}+b_i)/2$. Integration around the cell midpoint gives
\[
|e_i|\leq\frac{\eta\ell_i}{4(1-\eta)}.
\]
Boundary indifference is $2(b_i+\mu)=q_i(f)+q_{i+1}(f)$; thus
\[
\ell_{i+1}-\ell_i=4\mu-2(e_i+e_{i+1})
\geq4L-\frac{\eta}{2(1-\eta)}>2L
\]
for sufficiently small $\eta$. Since $\ell_1>0$ and $\sum_i\ell_i=1$, every feasible $j\geq2$ satisfies $1>L j(j-1)$. Babbling has one cell, so this bounds \emph{all} feasible counts uniformly.

For the upper estimate, suppose $f_h\to1$ in $C^2$, $\mu_h\to\bar\mu\in[L,H]$, and select any feasible partition for each pair $(f_h,\mu_h)$. After passing to a subsequence, the cell count is a fixed $j$ and all boundaries converge to $0=\bar b_0\leq\bar b_1\leq\cdots\leq\bar b_j=1$. For positive-length limiting cells, the conditional means converge to their uniform midpoints. For a vanishing cell, its conditional mean lies between its endpoints and converges to their common limit. Consequently the boundary equations pass to the limit, including at zero-length cells, and the weak limiting lengths satisfy
\[
\bar\ell_{i+1}=\bar\ell_i+4\bar\mu,
\qquad i=1,\ldots,j-1.
\]
Because $\bar\mu\geq L>0$, only the first cell can have zero length. If it is positive, the limit is an ordinary feasible uniform $j$-cell equilibrium. If $\bar\ell_1=0$, summing $\bar\ell_i=4\bar\mu(i-1)$ gives $\bar\mu=1/[2j(j-1)]=\tau_j$; deleting that empty first cell leaves the uniform $(j-1)$-cell equilibrium. The normalized payoffs converge as well: the integral of squared error over an empty cell tends to zero, while the integrals over the other cells and $c\mu_h^2$ converge. At $\tau_j$ the degenerate $j$-cell value and the selected $(j-1)$-cell value coincide. Hence every such limiting payoff is at most $W^0(\bar\mu)$, including when $j=1$.

If \eqref{eq:upper-partition-control} failed, there would be an $\epsilon>0$, densities $f_h\to1$, posteriors $\mu_h\in[L,H]$, and feasible partitions whose payoffs exceed $W^0(\mu_h)+\epsilon$; an approximate maximizer suffices because $M_f$ is a supremum. Extract a convergent posterior subsequence and apply the preceding argument. Continuity of $W^0$, including at every threshold, makes the limiting payoff strictly above $W^0(\bar\mu)$, a contradiction.
\end{proof}

\begin{proposition}[Robustness to smooth type-density perturbations]\label{prop:density-robustness}
Fix a binary support $0<L<H$, a prior mean $\mu_0\in(L,H)$, and a finite $c\geq0$. Suppose the uniform-type game has a line $T^0$ that strictly supports the finest-partition value $W^0$ on $[L,H]$ except at interior contacts $x^0<y^0$, which belong strictly to the $n$- and $m$-message regimes, respectively, with $n>m\geq2$. Suppose the two contact branches have negative second derivatives and $\mu_0\in(x^0,y^0)$. Maintain independence of $\theta$ and $B$, binary $B$, and quadratic preferences, but let $\theta$ have a strictly positive probability density $f\in C^2([0,1])$.

There is an $\varepsilon>0$ such that whenever $\|f-1\|_{C^2}<\varepsilon$ the following hold within the monotone pure-strategy partition equilibrium class. Unique nearby $n$- and $m$-cell equilibria are defined around unique nearby contacts $x^f<y^f$; at those contacts they satisfy NITS strictly and every coarser equilibrium violates NITS. A common tangent $T^f$ majorizes the principal's normalized payoff from \emph{every} feasible partition equilibrium at every $\mu\in[L,H]$, with equality only for the $n$-cell equilibrium at $x^f$ and the $m$-cell equilibrium at $y^f$. Consequently, under NITS at the two contacts (and arbitrary feasible partition selection elsewhere), the unique optimal posterior-mean law has support $\{x^f,y^f\}$ and weights
\[
\Prob(\mu_z=x^f)=\frac{y^f-\mu_0}{y^f-x^f},
\qquad
\Prob(\mu_z=y^f)=\frac{\mu_0-x^f}{y^f-x^f}.
\]
It strictly beats both no information and full revelation. The conclusion also holds under either player's conditionally preferred or ex-ante Pareto-efficient monotone partition selection.
\end{proposition}

\begin{proof}
Write $0=b_0<b_1<\cdots<b_j=1$, $\ell_i=b_i-b_{i-1}$, and
\[
q_i(f)=\frac{\int_{b_{i-1}}^{b_i}\theta f(\theta)\,d\theta}
{\int_{b_{i-1}}^{b_i}f(\theta)\,d\theta}.
\]
The boundary indifference equations are
\begin{equation}
F_i(b,f,\mu)\equiv2(b_i+\mu)-q_i(f)-q_{i+1}(f)=0,
\qquad i=1,\ldots,j-1. \label{eq:perturbed-boundaries}
\end{equation}
For the uniform density, $q_i(1)=(b_{i-1}+b_i)/2$; the Jacobian of $F$ in the interior boundaries has diagonal $1$, neighboring entries $-1/2$, and eigenvalues $1-\cos(r\pi/j)>0$ for $r=1,\ldots,j-1$. The implicit-function theorem thus gives a unique locally smooth $j$-cell partition at every nondegenerate uniform one. Its communication payoff
\begin{equation}
V_j^f(\mu)=-\sum_{i=1}^j
\int_{b_{i-1}}^{b_i}(\theta-q_i(f))^2f(\theta)\,d\theta,
\qquad W_j^f(\mu)=V_j^f(\mu)+c\mu^2, \label{eq:perturbed-value}
\end{equation}
is locally $C^2$ in $\mu$ and converges in $C^2$ to the corresponding uniform-type branch as $f\to1$ in $C^2$.

Lemma~\ref{lem:partition-upper-control} now controls \emph{all} feasible partition payoffs uniformly, including partitions that approach a threshold and lose their first cell. It supplies the one-sided upper bound \eqref{eq:upper-partition-control} without requiring an off-path map of perturbed thresholds.

In small fixed neighborhoods of $x^0$ and $y^0$, the sequential compactness argument of Lemma~\ref{lem:partition-upper-control} and \eqref{eq:perturbed-boundaries} exclude any finer feasible partition: a weak limit of one would have to be feasible at a posterior strictly inside $I_n$ or $I_m$. For each coarser partition, the uniform-type continuation payoff is strictly below that of the appropriate contact branch, by the adjacent-equilibrium ranking. There are only finitely many cell counts, and the implicit-function theorem plus compactness excludes additional fixed-count solutions near either contact. Shrinking the neighborhoods and $\varepsilon$ if needed, we have $M_f=W_n^f$ throughout the first neighborhood and $M_f=W_m^f$ throughout the second. These branches remain strictly concave.

The NITS inequalities also persist \emph{without} an off-path threshold assumption. For a uniform $j$-cell partition the lowest type's criterion is $q_{1,j}(1)-2\mu\leq0$, with
\[
q_{1,j}(1)-2\mu=\frac1{2j}-(j+1)\mu.
\]
At $x^0\in(\tau_{n+1},\tau_n)$ this expression is strictly negative for $j=n$ and strictly positive for every $j<n$; at $y^0\in(\tau_{m+1},\tau_m)$ the analogous statements hold for $j=m$ and $j<m$. All relevant conditional means and partitions vary continuously, and finer partitions remain infeasible nearby. Hence NITS uniquely selects the indicated partition outcomes at the shifted contacts. The principal and expert also rank those outcomes highest: the expert's ex-ante conditional communication loss under a partition is $-V_j^f(\mu)+\mu^2$, since $\E_f[q_i(f)-\theta]=0$.

Finally, strict concavity and $C^2$ convergence of the two contact branches give a unique nearby common tangent $T^f$ by the nonsingular contact-system Jacobian in the proof of Proposition~\ref{prop:geometric-robustness}. In the two contact neighborhoods, $T^f-W_n^f$ and $T^f-W_m^f$ vanish only at their respective contact points and are strictly positive elsewhere. On the compact complement the original gap $T^0-W^0$ has a positive minimum; \eqref{eq:upper-partition-control} and uniform convergence of $T^f$ preserve strict domination of every partition payoff. Thus $T^f$ is globally supporting with exactly the asserted equality set. Binary Bayes plausibility implements the two-contact law, and affinity plus the equality set proves uniqueness and both strict welfare comparisons. The two contacts remain in their respective $n$- and $m$-message regimes, while the corresponding off-path feasibility boundaries may move; when $n\geq m+2$, the regimes remain nonadjacent.
\end{proof}

The proposition is local: it supplies no numerical perturbation radius, especially for a support endpoint with a small gap. It keeps the expert's type independent of the residual state and does not cover atoms, nonquadratic preferences, unrestricted mixed equilibria, or arbitrary on-path babbling. It proves contact-level NITS and global domination of all feasible monotone partition payoffs; it does not require a global case map of perturbed feasibility thresholds.

\section{Proofs for Section~\ref{sec:delegation}}\label{app:delegation-proofs}

\begin{proof}[Proof of Lemma~\ref{lem:delegation-value}]
\emph{Step 0: the two rules and their losses.} Write the constraint for $(\theta,\hat\theta)$ and for $(\hat\theta,\theta)$ and add the two inequalities. The squares cancel and leave $(b(\hat\theta)-b(\theta))(\hat\theta-\theta)\geq0$, so every feasible rule is nondecreasing.

Under the cap with $\mu\leq1/2$, the available actions are $[\mu,1-\mu]$. Each type picks the available action closest to her ideal $\theta+\mu$, which is $\min\{\theta+\mu,1-\mu\}$, so the cap is feasible. The principal loses $\mu^2$ on each type below $1-2\mu$ and $(1-\mu-\theta)^2$ on each type above it, for a total of $\mu^2(1-2\mu)+\frac23\mu^3=\mu^2-\frac43\mu^3$. The constant rule is feasible and loses $\int_0^1(\frac12-\theta)^2d\theta=\frac1{12}$. It remains to show that no feasible rule loses less.

\emph{Step 1: a rule is a menu.} Let $D$ be the set of actions the rule uses, together with its limit points. A type with ideal $y=\theta+\mu$ must receive the action in $D$ closest to $y$; otherwise she would gain by reporting the type that receives a closer action. Two closest actions exist only for countably many types (those at the midpoints of the gaps in $D$), which does not affect any payoff.

\emph{Step 2: a lower bound on the principal's loss.} Let $x$ be the action a type receives and $e\equiv x-y$ the expert's error relative to her own ideal. The principal's loss on that type is
\begin{equation}
(x-\theta)^2=(e+\mu)^2=\mu^2+e^2+2\mu e. \label{eq:loss-error-identity}
\end{equation}
A type who receives an action at or above her ideal ($e\geq0$) therefore costs the principal at least $\mu^2$. The principal can do better than $\mu^2$ only on types who are pushed \emph{below} their ideal.

Let $\kappa$ be the highest action in $D$ that does not exceed $1+\mu$, the highest ideal. If there is no such action, every type receives an action above her ideal, every type costs at least $\mu^2$, and the total loss is at least $\mu^2$, which is at least the bound \eqref{eq:delegation-loss-bound} below. Otherwise, split the types into three groups.
\begin{enumerate}[label=(\alph*),leftmargin=2.2em]
\item \emph{Ideals at or below $\kappa$.} A type whose ideal lies in $D$ gets it, so $e=0$. Every other type in this group has her ideal inside a gap $(x_1,x_2)$ of $D$, and she takes $x_1$ if her ideal is in the left half of the gap and $x_2$ if it is in the right half. Her error therefore falls from $0$ to $-g/2$ across the left half and from $+g/2$ to $0$ across the right half, where $g=x_2-x_1$. These errors cancel, so over a gap that lies entirely within the ideals $[\mu,\kappa]$ they sum to zero. Only the lowest gap can begin below $\mu$. The types it loses are its leftmost ones, all with negative errors, unless the cut passes the midpoint, in which case every remaining type has a positive error. Either way the remaining errors sum to a nonnegative number. (If the lowest gap has no lower end, every type in it receives an action above her ideal.) Summing over all gaps, the errors in group (a) add up to at least zero, so by \eqref{eq:loss-error-identity} the group's total loss is at least $\mu^2$ times its size.
\item \emph{Ideals above $\kappa$, receiving the next action above $\kappa$.} These types receive more than their ideal, so each costs at least $\mu^2$.
\item \emph{Ideals above $\kappa$, receiving $\kappa$.} Each costs exactly $(\kappa-\theta)^2$.
\end{enumerate}
Each type thus contributes, on average within its group, at least the smaller of $\mu^2$ and $(\kappa-\theta)^2$. Hence
\begin{equation}
\int_0^1\bigl(b(\theta)-\theta\bigr)^2d\theta\ \geq\ \int_0^1\min\bigl\{(\kappa-\theta)^2,\ \mu^2\bigr\}\,d\theta. \label{eq:delegation-loss-bound}
\end{equation}
In words: at best, each type costs the principal either the full bias $\mu^2$ or the loss from a single common action $\kappa$, whichever is smaller.

\emph{Step 3: the best value of $\kappa$.} Let $q(u)=\min\{u^2,\mu^2\}$. It is symmetric around zero and rises with $|u|$. The right side of \eqref{eq:delegation-loss-bound} equals $\int_{-\kappa}^{1-\kappa}q(u)\,du$, the integral of $q$ over an interval of length one. Its derivative with respect to $\kappa$ is $q(\kappa)-q(1-\kappa)$. This is nonpositive for $\kappa<1/2$ and nonnegative for $\kappa>1/2$, so the bound is smallest at $\kappa=1/2$, where it equals
\[
\int_{-1/2}^{1/2}\min\{u^2,\mu^2\}\,du=
\begin{cases}
\mu^2-\frac43\mu^3, & \mu\leq\frac12,\\[4pt]
\frac1{12}, & \mu\geq\frac12.
\end{cases}
\]
These are exactly the losses of the cap and of the constant rule in Step~0.
\end{proof}

\begin{proof}[Proof of Proposition~\ref{prop:delegation-design}]
We use the same argument as in Theorem~\ref{thm:global-skipping}. If a straight line lies above $W^D$ on $[L,H]$ and touches it only on a set $E$, then no study can beat the value of the line at $\mu_0$, and a study attains that value exactly when all its posteriors lie in $E$. Write $f=W^D$. We also use two elementary facts.
\begin{itemize}[leftmargin=2.2em]
\item A strictly concave function lies strictly below each of its tangents, except at the point of tangency.
\item A convex function lies below the straight line joining any two points on its graph, and strictly below it in between unless the function is linear there. Likewise, if it lies weakly below a line at two points and strictly below at one of them, it lies strictly below the line in between.
\end{itemize}

\emph{(a)} Here $f$ is strictly concave on $[L,H]$. The tangent at $\mu_0$ touches $f$ only at $\mu_0$, so revealing nothing is uniquely optimal.

For (b) and (c), let $H>\hat\mu$. On the concave part, the tangent gap $\varphi$ is strictly decreasing, because $\varphi'(x)=f''(x)(H-x)<0$. At $x=\max\{\hat\mu,L\}$ it is negative: on the convex part $f$ is not linear, so $f(H)$ lies strictly above the tangent there.

\emph{(b)} Suppose $\varphi(L)\leq0$: the tangent at $L$ passes weakly below $f(H)$. Take the line through $(L,f(L))$ and $(H,f(H))$. On the concave part $f$ lies below its tangent at $L$, which lies below this line, so $f$ is strictly below the line there except at $L$. On the convex part $f$ is strictly below the line except where the line meets it, at $H$ (and at $L$ if there is no concave part). The line touches $f$ only at $L$ and $H$, so full revelation is uniquely optimal. When $L\geq\hat\mu$ there is no concave part, and $\varphi(L)\leq0$ holds automatically; this covers $c\geq1$.

\emph{(c)} Suppose $\varphi(L)>0$. Then $L<\hat\mu$ by the last remark, and since $\varphi$ falls from a positive to a negative value on $[L,\hat\mu]$, it has a unique zero $x^D$ there. The tangent at $x^D$ passes through $(H,f(H))$. It lies strictly above $f$ on the concave part except at $x^D$, and strictly above $f$ on the convex part except at $H$. It therefore touches $f$ only at $x^D$ and $H$, which gives \eqref{eq:delegation-weights} for $\mu_0>x^D$. For $\mu_0\leq x^D$, the tangent at $\mu_0$ has $\varphi(\mu_0)\geq0$, so it passes weakly above $f(H)$. It lies strictly above $f$ everywhere except at $\mu_0$ (and also at $H$ when $\mu_0=x^D$). In both cases the only posterior distribution with mean $\mu_0$ on the touching set is the point mass at $\mu_0$, so revealing nothing is uniquely optimal.

\emph{Closed forms.} For $x\leq H\leq1/2$, $f$ is a cubic on $[x,H]$, so its Taylor expansion around $x$ is exact:
\[
f(H)-f(x)-f'(x)(H-x)=\tfrac12f''(x)(H-x)^2+\tfrac16f'''(H-x)^3
=(H-x)^2\left[\tfrac83x+\tfrac43H-(1-c)\right],
\]
using $f''(x)=-2(1-c)+8x$ and $f'''=8$. This gives \eqref{eq:xD-closed-form}. For $H>1/2$, the value $f(H)=-\frac1{12}+cH^2$ falls short of the cubic formula at $H$ by exactly $\frac{(2H-1)^2(4H+1)}{12}$, which gives \eqref{eq:phi-closed-form}.
\end{proof}


\clearpage
\addcontentsline{toc}{section}{References}
\begin{thebibliography}{99}

\bibitem[Agastya, Bag, and Chakraborty(2014)]{AgastyaBagChakraborty2014}
Agastya, Murali, Parimal Kanti Bag, and Indranil Chakraborty. 2014.
``Communication and Authority with a Partially Informed Expert.''
\emph{RAND Journal of Economics} 45(1): 176--197.

\bibitem[Alonso and Matouschek(2008)]{AlonsoMatouschek2008}
Alonso, Ricardo, and Niko Matouschek. 2008.
``Optimal Delegation.''
\emph{Review of Economic Studies} 75(1): 259--293.

\bibitem[Ambrus and Zhao(2026)]{AmbrusZhao2026}
Ambrus, Attila, and Luke Zhao. 2026.
``Optimal Information Structures for Strategic Communication.''
Working paper, Duke University. An earlier version circulated as ``Optimal Information Structures with Information Control'' (2024).

\bibitem[Anti\'c and Persico(2020)]{AnticPersico2020}
Anti\'c, Nemanja, and Nicola Persico. 2020.
``Cheap Talk with Endogenous Conflict of Interest.''
\emph{Econometrica} 88(6): 2663--2695.

\bibitem[Arieli, Geffner, and Tennenholtz(2025)]{ArieliGeffnerTennenholtz2025}
Arieli, Itai, Ivan Geffner, and Moshe Tennenholtz. 2025.
``Optimal Information Design in Sender-Receiver Cheap Talk Interactions.''
\emph{Electronic Proceedings in Theoretical Computer Science} 437: 31--45.

\bibitem[Beetsma and Debrun(2016)]{BeetsmaDebrun2016}
Beetsma, Roel, and Xavier Debrun. 2016.
``Fiscal Councils: Rationale and Effectiveness.''
IMF Working Paper 16/86, International Monetary Fund.

\bibitem[Bergemann and Morris(2019)]{BergemannMorris2019}
Bergemann, Dirk, and Stephen Morris. 2019.
``Information Design: A Unified Perspective.''
\emph{Journal of Economic Literature} 57(1): 44--95.

\bibitem[Chen(2012)]{Chen2012}
Chen, Ying. 2012.
``Value of Public Information in Sender--Receiver Games.''
\emph{Economics Letters} 114(3): 343--345.

\bibitem[Chen, Kartik, and Sobel(2008)]{CKS2008}
Chen, Ying, Navin Kartik, and Joel Sobel. 2008.
``Selecting Cheap-Talk Equilibria.''
\emph{Econometrica} 76(1): 117--136.

\bibitem[Crawford and Sobel(1982)]{CrawfordSobel1982}
Crawford, Vincent P., and Joel Sobel. 1982.
``Strategic Information Transmission.''
\emph{Econometrica} 50(6): 1431--1451.

\bibitem[Deimen and Szalay(2019)]{DeimenSzalay2019AER}
Deimen, Inga, and Dezs\H{o} Szalay. 2019.
``Delegated Expertise, Authority, and Communication.''
\emph{American Economic Review} 109(4): 1349--1374.

\bibitem[Fischer and Stocken(2001)]{FischerStocken2001}
Fischer, Paul E., and Phillip C. Stocken. 2001.
``Imperfect Information and Credible Communication.''
\emph{Journal of Accounting Research} 39(1): 119--134.

\bibitem[Frones(2026)]{Frones2026}
Frones, Luis. 2026.
``Speak or Wait: Strategic Delay and the Value of Public Information.''
\emph{Economics Letters} 258: 112752.

\bibitem[Gilligan and Krehbiel(1987)]{GilliganKrehbiel1987}
Gilligan, Thomas W., and Keith Krehbiel. 1987.
``Collective Decision-Making and Standing Committees: An Informational Rationale for Restrictive Amendment Procedures.''
\emph{Journal of Law, Economics, and Organization} 3(2): 287--335.

\bibitem[Goltsman et~al.(2009)]{GoltsmanEtAl2009}
Goltsman, Maria, Johannes H\"orner, Gregory Pavlov, and Francesco Squintani. 2009.
``Mediation, Arbitration and Negotiation.''
\emph{Journal of Economic Theory} 144(4): 1397--1420.

\bibitem[Holmstr\"om(1977)]{Holmstrom1977}
Holmstr\"om, Bengt. 1977.
``On Incentives and Control in Organizations.''
PhD dissertation, Stanford University.

\bibitem[Holmstr\"om(1984)]{Holmstrom1984}
Holmstr\"om, Bengt. 1984.
``On the Theory of Delegation.''
In \emph{Bayesian Models in Economic Theory}, edited by Marcel Boyer and Richard E. Kihlstrom, 115--141. Amsterdam: North-Holland.

\bibitem[Ivanov(2010)]{Ivanov2010}
Ivanov, Maxim. 2010.
``Informational Control and Organizational Design.''
\emph{Journal of Economic Theory} 145(2): 721--751.

\bibitem[Ivanov and Sam(2022)]{IvanovSam2022}
Ivanov, Maxim, and Alex Sam. 2022.
``Cheap Talk with Private Signal Structures.''
\emph{Games and Economic Behavior} 132: 288--304.

\bibitem[Jain(2018)]{Jain2018}
Jain, Vasudha. 2018.
``Bayesian Persuasion with Cheap Talk.''
\emph{Economics Letters} 170: 91--95.

\bibitem[Kamenica and Gentzkow(2011)]{KamenicaGentzkow2011}
Kamenica, Emir, and Matthew Gentzkow. 2011.
``Bayesian Persuasion.''
\emph{American Economic Review} 101(6): 2590--2615.

\bibitem[Kr\"ahmer(2021)]{Krahmer2021}
Kr\"ahmer, Daniel. 2021.
``Information Design and Strategic Communication.''
\emph{American Economic Review: Insights} 3(1): 51--66.

\bibitem[Kreutzkamp and Lou(2025)]{KreutzkampLou2025}
Kreutzkamp, Sophie, and Yichuan Lou. 2025.
``Persuasion without Ex-Post Commitment.''
\emph{Journal of Economic Theory} 228: 106058.

\bibitem[Li(2004)]{Li2004}
Li, Ming. 2004.
``To Disclose or Not to Disclose: Cheap Talk with Uncertain Biases.''
Working Paper 04003, Concordia University.

\bibitem[Li and Madar\'asz(2008)]{LiMadarasz2008}
Li, Ming, and Krist\'of Madar\'asz. 2008.
``When Mandatory Disclosure Hurts: Expert Advice and Conflicting Interests.''
\emph{Journal of Economic Theory} 139(1): 47--74.

\bibitem[Lyu and Suen(2026)]{LyuSuen2026}
Lyu, Qianjun, and Wing Suen. 2026.
``Information Design in Cheap Talk.''
\emph{Economic Theory}, published online October 2026.

\bibitem[Melumad and Shibano(1991)]{MelumadShibano1991}
Melumad, Nahum D., and Toshiyuki Shibano. 1991.
``Communication in Settings with No Transfers.''
\emph{RAND Journal of Economics} 22(2): 173--198.

\bibitem[Miyahara and Sadakane(2020)]{MiyaharaSadakane2020}
Miyahara, Yasuyuki, and Hitoshi Sadakane. 2020.
``Communication Enhancement through Information Acquisition by Uninformed Player.''
KIER Discussion Paper 1050, Kyoto University.

\bibitem[Moreno de Barreda(2024)]{MorenoDeBarreda2024}
Moreno de Barreda, In\'es. 2024.
``Cheap Talk with Two-Sided Private Information.''
\emph{Games and Economic Behavior} 148: 97--118.

\bibitem[Morgan and Stocken(2003)]{MorganStocken2003}
Morgan, John, and Phillip C. Stocken. 2003.
``An Analysis of Stock Recommendations.''
\emph{RAND Journal of Economics} 34(1): 183--203.

\bibitem[Niskanen(1971)]{Niskanen1971}
Niskanen, William A. 1971.
\emph{Bureaucracy and Representative Government}.
Chicago: Aldine-Atherton.

\end{thebibliography}
\end{document}